\documentclass[%
 reprint,
showpacs,preprintnumbers,
 amsmath,amssymb,
 aps,
]{revtex4-1}

\usepackage{graphicx}
\usepackage{dcolumn}
\usepackage{bm}
\usepackage{amssymb} 

\usepackage[pagebackref=false,colorlinks,linkcolor=blue,filecolor=green,urlcolor=blue ,citecolor=blue]{hyperref}
\usepackage{ulem}
\usepackage{placeins}
\usepackage{float}
\usepackage{comment}
\usepackage{soul}
\usepackage{xcolor}

\begin{document}

\title{Spin responses of a disordered helical superconducting edge under a Zeeman field}
\author{Zeinab Bakhshipour and Mir Vahid Hosseini}
 \email[Corresponding author: ]{mv.hosseini@znu.ac.ir}
\affiliation{Department of Physics, Faculty of Science, University of Zanjan, Zanjan 45371-38791, Iran}

\begin{abstract}

We investigate analytically and numerically the effects of disorder on the helical edge of a 2D topological insulator in the presence of a Zeeman field and superconductivity. Employing bosonization and a renormalization-group analysis, we study how impurity potentials modify charge- and spin-density wave correlations as well as superconducting pair correlations. Our results reveal that the Zeeman field controls the competition: in the attractive regime, it amplifies the superconducting gap, while in the repulsive regime, it stabilizes impurity effects by keeping the system longer in the relevant regime for disorder. We also find that disorder induces logarithmic suppression of density-wave correlations, while at the same time introducing positive logarithmic corrections that enhance superconducting pairing correlations and contribute to their stability. These effects directly modify the scaling of spin conductance, providing experimentally accessible signatures of the interplay between disorder and superconductivity in topological edge channels.
\end{abstract}

\maketitle

\section {Introduction} \label{s1}

One-dimensional helical liquids, realized, for instance, at the edges of two-dimensional topological insulators~\cite{Kane2005,Konig2007,Wu2006,Bernevig2006,Hassan2010,Sinova2015,Hsu2021,Hsu2025} and  their constructions like helical anti-wires~\cite{Fleckenstein2021} and quantum point contacts~\cite{Heedt2017,Fleckenstein2018a,Strunz2020}, offer a unique candidate to study spin transport~\cite{Copenhaver2022} in strongly correlated quantum matter. The fate of current- and spin-carrying elementary excitations~\cite{Giamarchi1989} in these systems has attracted significant attention in recent years due to their deep connection with static~\cite{Bahari2016,Bahari2019,Jangjan2021} and dynamical~\cite{Jangjan2022} topological phases~\cite{Qi2011} and potential applications in spintronics and quantum computation. In addition, helical states induced by Rashba spin-orbit coupling~\cite{Sun2007,Manchon2015,Fleckenstein2018b}, nuclear order~\cite{Braunecker2009a,Braunecker2009b,Braunecker2010,Braunecker2015,Hsu2017,Hsu2018} or hinge~\cite{Hsu2018b} are recognized as leading platforms for hosting Majorana fermions~\cite{Kitaev2001,Alicea2012,Beenakker2013,Sato2017}. Such systems, due to the interplay between the superconducting proximity effect and the magnetic field~\cite{Lutchyn2010} or even without the magnetic field~\cite{Orth2015,Pedder2017} form topological superconducting wires. 

The effects of interactions in topological superconducting wires that support Majorana fermions have been investigated~\cite{Stoudenmire2011}. It has been showed that the interactions not only suppress the bulk gap ~\cite{Gangadharaiah2011,Sela2011}, but also lead to the accessibility of Majorana fermions and topological superconductors in a weaker magnetic field and on a wider range of chemical potential. This makes the phase immune to chemical potential changes caused by disorder in the wire.
An overview of the properties, fundamentals and realization of topological superconductors, Majorana fermions, and the role of spin-orbit coupling of these states is given in Ref. \cite{Sato2017}. One- and two-dimensional models and hybrid superconductor-semiconductor devices~\cite{Mourik2012,Prada2020,Agaee2023} were proposed that host non-Abelian Majorana zero modes \cite{Alicea2012,Sato2003} and parafermions \cite{Alicea2016}. The intrinsic robustness of topological superconducting platforms against local perturbations has motivated proposals for fault-tolerant, topologically protected quantum computation \cite{Sarma2015,Aasen2016,Yazdani2023}. The realization of such ideas became possible thanks to the discovery of topological insulators by providing platforms such as helical edge models. It has been proposed that pair interaction mediated by spin fluctuations as a joint thread can link a wide class of superconducting materials \cite{Scalapino2012}. Specifically, the close relation of the spin-density-wave phase to an s-wave superconductor with a change of sign can be a common feature.

The characteristic of helical edges is spin–momentum locking~\cite{Wu2006,Xu2006}  producing spin-resolved counter-propagating electrons along the same edge, forming the basis of their distinctive transport properties and forbids conventional elastic backscattering in the absence of time-reversal symmetry breaking. This property makes helical edges ideal candidates for spintronics~\cite{Datta1990,Zutic2004,Fabian2007} applications with dissipationless transport. However, realistic systems inevitably contain perturbations such as impurities in 1D~\cite{Giamarchi1988} and 2D~\cite{Pooyan2022,Heydari2025}, external magnetic fields~\cite{Gangadharaiah2008}, and proximity-induced superconductivity~\cite{Liu2013,Virtanek2012,Fidkowski2012,Braunecker2013b,Wang2021}. So they alter the idealized scenario in profound ways and strongly affect quantum correlation functions \cite{Rice1976,Bouchoule2025,Mahdavifar2015} and transport properties.

The effect of disorders in one dimension is very intriguing as a result of the limited available space. Even weak impurities can couple to helical degrees of freedom and generate backscattering channels \cite{Kane1992}, particularly in the presence of a driven Rashba impurity \cite{Privitera2020}. Magnetic impurities are especially detrimental, as they allow spin-flip processes that directly reduce the spin current~\cite{Tanaka2011}. When a helical edge is brought into contact with a conventional s-wave superconductor, the situation becomes even more intricate; Andreev reflection converts incoming electrons into opposite-spin holes, which, while enabling superconducting correlations~\cite{Fu2008}, simultaneously diminishes the net spin conductance. Disorder further complicates this picture by interfering with Andreev processes and enhancing scattering, leading to stronger suppression of spin transport~\cite{Schmidt2012}.

The introduction of a Zeeman field adds yet another layer of competition. Moreover, it reduces the effective Luttinger parameter $K$, extending the relevance of disorder down to lower energy scales in the renormalization group (RG) and keeping impurity scattering effective over a broader low energy window \cite{Kane1992,Braunecker2013a}. Simultaneous consideration of these three ingredients, i.e., superconductivity, disorder, and Zeeman-induced symmetry breaking~\cite{Wei2006,Eriksson2015}, provides a unique framework investigating mechanisms of nontrivial spin transport. Such a setting can reveal phase transitions between spin-conducting states, impurity-induced insulating regimes, and proximity-induced superconducting phases. Understanding this competition is not only of fundamental interest for correlated one-dimensional physics, but also of practical importance for designing topological devices that aim to exploit robust spin currents. Although previous studies have addressed the effects of a single impurity, disorder, or superconducting proximity individually~\cite{Kane1992,Tanaka2011,Fu2008}, the combined influence of all three factors remains largely unexplored.

In this work, we explore the interplay between disorder, Zeeman field, and proximity-induced superconductivity in one-dimensional helical Luttinger liquids. Our focus is on understanding how these perturbations affect spin transport along the helical edge. Specifically, we analyze how random impurities scattering modify the power-law behavior of spin conductance and competes with superconducting pairing, particularly under the influence of a Zeeman field that tunes the effective interaction parameter. We distinguish between the effects of single impurity and many-impurities disorder, uncovering how collective backscattering can lead to stronger suppression of spin conductance and, in some regimes, to localization. Moreover, we study how superconducting correlations, reinforced by attractive interactions, may counteract the disorder-induced degradation of transport, resulting in a rich interplay that governs the low-temperature behavior of both charge and spin currents. In addition, we find that disorder introduces a logarithmic enhancement in the pairing correlations, while it leads to a suppression of both charge- and spin-density-wave correlations.

The structure of this paper is organized as follows. In Sec. \ref{s2}, we introduce the model and theoretical framework of a partially spin-mixed helical superconductor under a Zeeman field. In Sec. \ref{s3}, we analyze the spin conductance in the presence of a single impurity using the RG approach. Section \ref{s4} explores the fate of spin transport in the presence of random impurities (disorder). In Sec. \ref{s5}, we make effective Hamiltonian with both Zeeman and superconductor relevant-gaps and investigate the effect of disorder on various correlation functions. Finally, in Sec. \ref{s6}, we present a summary of our main results. Also, some detailed calculations are included in Appendices.

\section {Model}\label{s2}
\begin{figure}[t!]
    \centering
    \includegraphics[width=8.5cm]{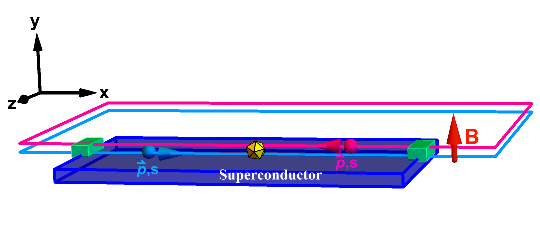}
   \caption{(Color online) Schematic of a superconducting PMH edge in the presence of insulator impurities. We look for the effect of a single impurity placed at $x=0$ represented in yellow color.}
    \label{fig1}
\end{figure}

The setup we considered is the edge of a quantum spin Hall insulator~\cite{Qi2011} with partially mixed helical (PMH) states due to the application of a magnetic field~\cite{Soori2012,Wozny2018,Hosseini2020}. We place this edge in contact with a conventional superconductor. The resulting system forms a one-dimensional superconducting PMH (super-PMH) state~\cite{Bakhshipour2025}. Unlike fully filtered helical spin states that are immune to impurities, especially magnetic impurities, they are now susceptible to strain against impurities. With this perspective, we visualize a more realistic super-helical system by including charge or spin impurity barriers (see Fig. \ref{fig1}). Then, considering the interactions in the system, we will have a strongly correlated example that can be described within the framework of Luttinger liquid theory~\cite{Giamarchi2004}. 

In the absence of electron-electron interactions, the Hamiltonian of the clean system is the sum of the Hamiltonians of the PMH state, $\mathcal{H}_{PMH}$, which is obtained by projecting the Zeeman‑perturbed helical edge spectrum onto the low‑energy modes near the Fermi points $\pm k_F$~\cite{Bakhshipour2024}, and the superconducting state, $\mathcal{H}_{sup}$, that is,
$\mathcal{H}_{sup-PMH}=\mathcal{H}_{PMH}+\mathcal{H}_{sup}$ with
\begin{align}
    \mathcal{H}_{PMH}=\psi^{\prime\dagger}\tilde{v}_F k_x\sigma^x\psi^\prime,\\\mathcal{H}_{sup}=\Delta_s\psi^\prime_R\psi^\prime_L+h.c.,
\end{align}
where $\tilde{v}_F=\hbar v_F\sqrt{1-\frac{\Delta_z^2}{\epsilon_F^2}}$ is the magnetized Fermi velocity with $v_F$, $\epsilon_F$, and $\Delta_z$ are the Fermi velocity, the Fermi energy, and the Zeeman gap, respectively. In addition, $\Delta_s$ is superconducting gap. The fermionic field operator is linearized around the Fermi points $\pm k_F$ and written as $\psi^\prime=e^{ik_F x}\psi^\prime_R+e^{-ik_F x}\psi^\prime_L$ with the chiral spinor fields $\psi^\prime_r=\chi_r(k_F)\psi_r$ where $\psi_r$ are scalar chiral fermionic operators and $\chi_r(k_F)=\frac{1}{\sqrt{2}}\left(\begin{matrix}re^{-i\vartheta_{k_F}}&e^{i\vartheta_{k_F}}\end{matrix}\right)^\dagger$ are the eigen states of the PMH edge with $\vartheta_{k_F}=\frac{1}{2}\arctan(\-\frac{\Delta_z}{v_F k_F})$.

We also include electron-electron interactions with Hamiltonian ~\cite{Bakhshipour2024},
\begin{align}
\mathcal{H}_{int} = &\, g_1 \, \psi_L^{\prime\dagger}(x) \psi_R^{\prime\dagger}(x) \psi_L^\prime(x) \psi_R^\prime(x) \nonumber \\
+ &\, g_2 \, \psi_L^{\prime\dagger}(x) \psi_L^\prime(x) \psi_R^{\prime\dagger}(x) \psi_R^\prime(x) \nonumber \\
+ &\, \frac{g_4}{2} \sum_{r=R,L} [\psi_r^{\prime\dagger}(x) \psi_r^\prime(x)]^2,
\end{align}
where $g_1$, $g_2$, and $g_4$ denote the backscattering, dispersive scattering, and forward scattering interaction strengths, respectively. We employ the standard bosonization identities for the scalar chiral fermionic fields as $\psi^{(\dagger)}_r=\frac{1}{\sqrt{2\pi a_0}}\eta_r e^{(-) ir\sqrt{4\pi}\phi_r(x)}$ with $a_0$ is a short-distance cutoff, $\eta_r$ are Klein factors ensuring proper
anticommutation relations between different chiral branches, and 
$\phi_r(x)$ are chiral bosonic fields. In the bosonized language, the total Hamiltonian of the system takes the form~\cite{Bakhshipour2025} \begin{equation}
\begin{split}
\mathcal{H}_{PMH}=\frac{v}{2}\left[\frac{1}{K}(\partial_x\Phi)^2+K(\partial_x\Theta)^2\right],
\label{PMH}
\end{split}
\end{equation}
\begin{equation}
\begin{split}
\mathcal{H}_{sup}=\frac{\Delta}{\pi a_0}\cos(\sqrt{4\pi}\Theta),
\label{sup}
\end{split}
\end{equation}
where $\Delta=\Delta_s\sin(2\vartheta_{k_F})$. Also, $\Theta=\phi_R-\phi_L$ and $\Phi=\phi_R+\phi_L$. In addition, $K$ and $v$ are the magnetized versions of the Luttinger parameter and the velocity of the collective mode, given by
\begin{equation}
K=\sqrt{\frac{\tilde{v}_F -\frac{g_{fb}}{8\pi}+\frac{g_4}{8\pi}}{\tilde{v}_F +\frac{g_{fb}}{8\pi}+\frac{g_4}{8\pi}}},
\label{Luttingerparameter}
\end{equation}
and
\begin{equation}
v=\sqrt{(\tilde{v}_F +\frac{g_4}{8\pi})^2-(\frac{g_{fb}}{8\pi})^2},
\label{Luttingervelocity}
\end{equation}
where $g_{fb}=g_2-g_1$. In the following, we take $\hbar v_F/a=1$ as the unit of energy, the lattice constant $a=1$ as the length unit, and the Boltzmann constant $k_B=1$.

It should be noted that we effectively considered the Zeeman field effect at the edge of the PMH as a result of fermionic single-particle fields. Recall that in the representation $\vartheta_{k_F}=0$, the Zeeman term exists independently with a sine-Gordon term and its associated flow. We used this flow despite the effective presence of Zeeman in the quadratic Hamiltonian (with $\vartheta_{k_F}\neq0$).

We consider either a single charge (c) or spin (s) impurity. The Hamiltonian of the charge and spin impurities is as
\begin{align}
    \mathcal{H}_{imp-c}&=\sum_{r,r^\prime}\int dx V_c \delta(x) \psi^{\prime\dagger}_r\psi^\prime_{r^\prime}, \\\mathcal{H}_{imp-s}^i&=\int dx V_{si} \delta(x) \mathcal{O}^i  \quad i=(x,y,z),
\end{align}
where $V_c$ is the strength of the charge impurity and $V_{si}$ is the i{\it th} component of the spin impurity. $\delta(x)$ indicates the Dirac delta function and $\mathcal{O}^i=\sum_{r,r^\prime}\psi^{\prime\dagger}_r\sigma^i\psi^\prime_{r^\prime}$.
The bosonic forms of these terms are found as
\begin{align}
    \mathcal{H}_{imp-c}&=\int dx V_c \delta(x)\{\frac{1}{\sqrt{4\pi}}\partial_x\Phi(x,\tau)\nonumber\\&+\frac{\cos(2\vartheta_{k_F})}{\sqrt{4\pi}}\partial_x\Theta(x,\tau)\nonumber\\&+\frac{\sin(2\vartheta_{k_F)}}{\pi a_{0}}\cos(\sqrt{4\pi}\Phi(x,\tau)+2k_{f}x)\},
 \label{impH-c}
 \\\mathcal{H}^{x}_{imp-s}&=\int dx V_{sx} \delta(x)\frac{\cos(2\vartheta_{k_F})}{\sqrt{4\pi}}\partial_x\Theta(x,\tau),
  \label{impH-sx}
  \\
 \mathcal{H}^{y}_{imp-s}&=\int dx V_{sy} \delta(x)\{\frac{\sin(2\vartheta_{k_F})}{\sqrt{4\pi}}\partial_x\Phi(x,\tau)\nonumber\\&+\frac{1}{\pi a_{0}}\cos(\sqrt{4\pi}\Phi(x,\tau)+2k_{f}x)\},
   \label{impH-sy}\\
 \mathcal{H}^{z}_{imp-s}&=-\int dx V_{sz} \delta(x)\frac{\cos(2\vartheta_{k_F})}{\pi a_{0}}\nonumber\\&\times\sin(\sqrt{4\pi}\Phi(x,\tau)+2k_{f}x).
 \label{impH-sz}
\end{align}
Without loss of generality, we consider the single impurity at position $x=0$ on the super-PMH edge, as shown in Fig. \ref{fig1}. Accordingly, the spatial integrals for terms without spatial derivatives can be carried out. 
Whereas for terms involving spatial derivatives the integration over $x$ should left, since these derivatives must remain explicit at the level of the Hamiltonian density.
 In this case, Eqs. (\ref{impH-c})-(\ref{impH-sz}) simplify as
\begin{align}
    \mathcal{H}_{imp-c}&=\int dx V_c \delta(x)\{\frac{1}{\sqrt{4\pi}}\partial_x\Phi(x,\tau)\nonumber\\&+\frac{\cos(2\vartheta_{k_F})}{\sqrt{4\pi}}\partial_x\Theta(x,\tau)\}\nonumber\\&+\frac{V_c \sin(2\vartheta_{k_F})}{\pi a_0}\cos(\sqrt{4\pi}\Phi(0,\tau))
    \label{imp-c},\\
    \mathcal{H}_{imp-s}^x&=\frac{V_{sx} \cos(2\vartheta_{k_F})}{\sqrt{4\pi}}\partial_x\Theta(0,\tau),
    \label{imp-x}\\
     \mathcal{H}_{imp-s}^y&=\int dx V_{sy} \delta(x)\frac{\sin(2\vartheta_{k_F})}{\sqrt{4\pi}}\partial_x\Phi(x,\tau)\nonumber\\&+\frac{V_{sy}}{\pi a_0}\cos(\sqrt{4\pi}\Phi(0,\tau)),
    \label{imp-y}\\
     \mathcal{H}_{imp-s}^z&=\frac{V_{sz} \cos(2\vartheta_{k_F})}{\pi a_0}\sin(\sqrt{4\pi}\Phi(0,\tau)).
    \label{imp-z}
\end{align}
One observes that within the PMH state in the x-direction, a single spin impurity aligned with the x-axis does not create a gap in the system. We note that this impurity term corresponds to forward scattering and is therefore linear in the bosonic field. In the bosonized description such a term can indeed be absorbed into the quadratic part of the super-PMH Hamiltonian with a change in the superconducting coefficient in $\mathcal{H}_{sup}$. Consequently, it does not modify scaling dimensions or long-distance correlation exponents and is therefore not essential for the RG analysis. In contrast, because of the effect of the forward part of the charge impurity ($\partial_x\Phi$) in corrections, we keep it. 

It is worthwhile noting that other spin-mixing mechanisms (for instance random Rashba coupling \cite{Crepin}) can also produce interaction-assisted backscattering in helical edges. But there is a difference between the Zeeman field and disorder originating from random Rashba spin–orbit coupling. The Zeeman term directly breaks time-reversal symmetry and therefore permits  single-particle elastic backscattering at the noninteracting level. By contrast, random Rashba coupling preserves global time-reversal symmetry and generally yields spin-mixing operators whose ability to produce elastic backscattering relies on interaction- or higher-order processes (see, e.g., Crepin et al. \cite{Crepin} and Geissler et al. \cite{Geissler}). Thus, while both mechanisms provide routes to backscattering, the two belong to different symmetry classes and lead to quantitatively and sometimes qualitatively different low-energy behavior. In particular, when a Zeeman field and Rashba disorder are both present, the former perturbation is expected to dominate the low-energy transport and correlation physics and typically drives stronger RG flows than the latter one.

 \section{single impurity}\label{s3}


\subsection{Renormalization Group analysis}\label{s3-A}

At first, in addition to the gradient terms that represent a forward scattering, backward scattering terms appear in the form of sine-Gordon gapped fields, such as the superconducting term. With a more detailed analysis, we study the effects of the perturbation caused by the aforementioned barriers in addition to analyzing the effect of the interactions on all gapped fields. Based on the correlation functions method \cite{Giamarchi2004}, the RG analysis of pairing term in the PMH edge gives the flow equations as (see Appendix \ref{App-A})
\begin{align}
\frac{d\mathcal{Y}(l)}{dl}&=(2-K^{-1}(l))\mathcal{Y}(l),
\label{YRG}
\\
\frac{dK^{-1}(l)}{dl}&=-\frac{1}{4}\mathcal{Y}^2(l) K^{-2}(l)
\label{KRG},
\end{align}
where $\mathcal{Y}=\frac{4\Delta a_0}{v}$ is the dimensionless gap contact with $\Delta=\Delta_s\sin(2\vartheta_{k_F})$. Note that the effects of Zeeman gap are absorbed into the Luttinger parameters and $\vartheta_{k_F}$. Alternatively, one may treat the Zeeman and superconducting terms as two independent sine-Gordon perturbations of a helical edge. Following Ref. \cite{Bakhshipour2025} and the discussion at the end of Appendix \ref{App-A}, their RG flow equations are $\frac{d\mathcal{Y}_z(l)}{dl} = (2 - K(l)) \mathcal{Y}_z(l)$ and $\frac{d\mathcal{Y}_s(l)}{dl} = \left(2 - \frac{1}{K(l)}\right) \mathcal{Y}_s(l)$, where $\mathcal{Y}_z=\frac{\sqrt{2}\Delta_z a_0}{v}$ and $\mathcal{Y}_s=\frac{\sqrt{2}\Delta_s a_0}{v}$ are the dimensionless Zeeman and superconducting gap parameters, respectively. Each of these perturbations opens a gap into the helical edge. Thus, alongside the main procedure, i.e., the PMH-projected description, one can use these flows to examine the individual behavior of the Zeeman and superconducting couplings.

In addition, the flow equations for impurities are obtained by a scaling calculation as
\begin{equation}
\frac{dV_i(l)}{dl}=(1-K)V_i(l)  \quad  i=(c,sy,sz).
    \label{impRG}
\end{equation}
We note that the spin impurity strength in the x-direction is not renormalized because of the lack of power-law behavior. According to Eq. (\ref{YRG}), for $K>1/2$ the superconductivity coefficient enters its relevant regime and leads to a gap. Moreover, according to Eq. (\ref{Luttingerparameter}), in the presence of an electron repulsive interaction in the range $0<g_{fb}<1/2$, the magnetized Luttinger parameter goes beyond the boundary of $1/2$ and the relevant gap regime of superconductivity will emerge. In this case, if we increase the Zeeman field strength, $K$ decreases and the system enters the relevant regime for interaction values smaller than $1/2$. In contrast, when the interactions are attractive, increasing the Zeeman value makes the superconducting gap stronger, so that even weaker attractive interactions can create a stable superconducting gap.

\begin{figure}[t!]
    \centering
    \includegraphics[width=8cm]{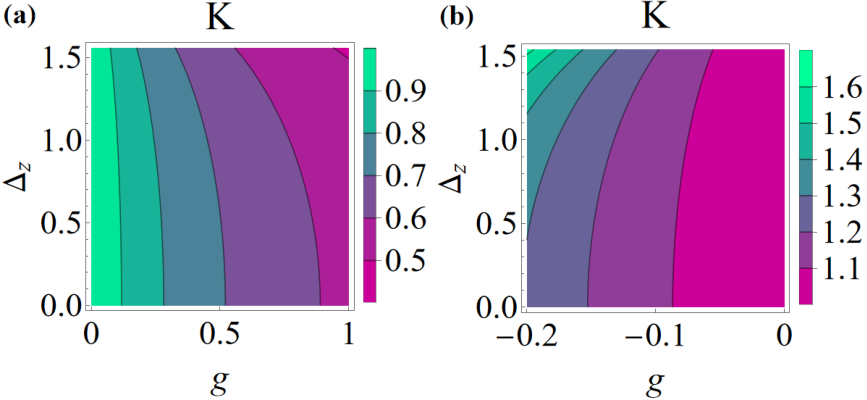}
   \caption{(Color online) Luttinger parameter as function the Zeeman gap and interactions in (a) the repulsive and (b) the attractive regimes. Here $g=\frac{g_{fb}}{8\pi}$ indicates the interaction effect.}
    \label{int2}
\end{figure}

Equation (\ref{impRG}) indicates that the impurity backscattering coefficient becomes relevant for $K < 1$. In the absence of a Zeeman field, electron–electron repulsion therefore plays a crucial role in generating an impurity‑induced pinning low-energy scale \cite{Giamarchi2004}. When the Zeeman field is turned on, the Luttinger parameter $K$ decreases from its Zeeman‑free value (see Fig. \ref{int2}(a)), driving the system deeper into the relevant regime of impurity backscattering. In other words, the Zeeman field enhances the impurity‑induced pinning effect in the presence of repulsive interactions. By contrast, for attractive interactions the Zeeman field increases $K$ (see Fig. \ref{int2}(b)), causing the impurity perturbation to flow from the relevant regime to the irrelevant regime under renormalization.

\begin{figure}[htbp]
    \centering
    \includegraphics[width=8cm]{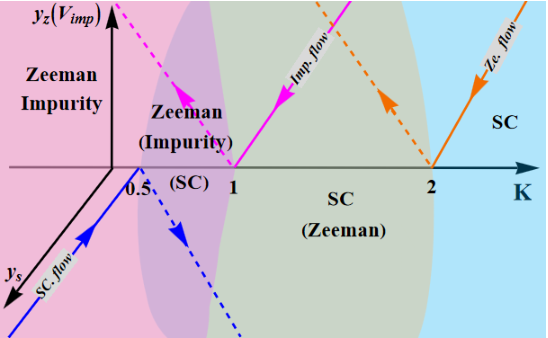}
   \caption{(Color online) RG flow of the superconducting (SC) ($\mathcal{Y}_{s}$, blue), impurity (Imp) ($V_{\mathrm{imp}}$, purple), and Zeeman (Ze) ($\mathcal{Y}_z$, orange) couplings as a function of the Luttinger parameter $K$. The vertical axes represent the corresponding dimensionless couplings strength under RG flow. Depending on the value of $K$, each perturbation may be relevant, marginal, or irrelevant, leading to distinct regions characterized by competing instabilities. The background colors indicate the dominant phase in each regime, while the phase listed in parentheses denotes the subdominant instability.
}
    \label{IZS-flow}
\end{figure}

Figure \ref{IZS-flow} shows the RG flow of the three competing perturbations, obtained from equations for the Zeeman and superconducting terms (discussed above) together with the impurity flow equation in Eq. (\ref{impRG}). The superconducting coupling $\mathcal{Y}_{s}$ (blue line), the impurity coupling $V_{{imp}}$ (purple line), and the Zeeman coupling $\mathcal{Y}_z$ (orange line) are plotted as functions of the Luttinger parameter $K$. The vertical axes represent the corresponding dimensionless couplings strength under RG flow. The slope of each curve reflects the scaling dimension of the associated operator. It determines whether the perturbation is relevant (growing under RG), marginal, or irrelevant (decreasing under RG). The background colors indicate the dominant phase in each parameter region, as determined by the most relevant coupling. While the subdominant channel is shown in parentheses.

In the figure, three boundary points $K$ with values of $1/2$, $1$, and $2$ are identified for superconductivity, impurity, and Zeeman, respectively. At $K<1/2$, where superconductivity is in an irrelevant regime, impurity and Zeeman tend to a strong coupling. This implies the enhanced role of backscattering in the repulsive regime. At $1/2<K<1$, despite the strong Zeeman, the superconductivity coefficient starts to gap out spin states and competes with the gap due to impurity. In the region $1<K<2$, the impurity loses its gapping ability. In this case, an intermediate superconducting gap coexists with a weaker Zeeman one. At $K>2$, the superconductivity gap is heading towards the strong coupling regime, where no traces of magnetic and impurity phases are observed. 

\subsection{Spin transport in $\bf{1/2<K<1}$ regime}\label{s4}

Spin transport characteristic in one-dimensional systems, particularly in helical Luttinger liquids, mutually affects the charge one. That is, the spin response to the external spin potential measuring by conductivity. In addition, in these systems, spin-momentum locking gives rise to protected transport channels that, in the ideal case, are immune to backscattering.

However, the presence of symmetry-breaking perturbations such as local impurities, an external Zeeman field, and proximity to a conventional s-wave superconductor introduces complex interactions that strongly affect correlation functions and transport properties. Impurities can induce local excitations, backscattering processes, and even drive transitions between conducting and insulating regimes. The Zeeman field, by breaking time-reversal symmetry, allows coupling between the right- and left-moving channels, thereby modifying the spin dynamics. On the other hand, superconducting proximity induces pairing correlations and nonlocal interactions at the edges, which may lead to the emergence of topological or trivial superconducting phases depending on the strength of the induced pairing. So, one expects that the resulting interplay between superconductivity, disorder, and Zeeman-induced symmetry breaking would create a delicate balance determining the fate of spin transport in such systems.

In the presence of a single impurity at position x=0, we should look for carriers transport around that point. So we examine (see Appendix \ref{App-B}) the edge resistance in the limit $L\sim0$~\cite{Giamarchi2004} with $R_{edge}=R_{sup-PMH}+R_{imp}$. Spin conductance and its impurity-induced  correction (see Appendix \ref{App-C}) are given by
\begin{equation}
    G_{s}\approx1-R_{edge}.
\end{equation}
 In the next two subsections, we calculate each part separately.

\subsection{Spin conductance of clean super-PMH states}
The resistance of the clean super-PMH state is defined by
\begin{equation}
    R_{\mathrm{sup\text{-}PMH}}
    = \left( \frac{1}{L} \sum_{q} \sigma_{\mathrm{sup\text{-}PMH}}(q) \right)^{-1},
\end{equation}
where $\sigma_{\mathrm{sup\text{-}PMH}}$ denotes the spin response of the superconducting proximity‑induced PMH edge. It is defined within linear‑response theory as the low‑frequency $(\omega \to 0)$ response of the edge spin current to a  Zeeman perturbation, whose explicit derivation is presented in Appendix \ref{App-B}. $L$ is the edge length. After evaluating the above relation (see Appendix~\ref{App-B}), $R_{sup-PMH}$ is obtained as 
\begin{align}
     R_{sup-PMH}&=\mathcal{Y}^2 \nonumber\\&\times  f(K^{-1})T^{2K^{-1}-3}\Big(1+\frac{g(K^{-1})}{T^2}\Big).
\end{align}
Here, T is the temperature.  $f(K^{-1})$ and $g(K^{-1})$ are functions of the luttinger parameter $K$. Their expressions are given in Appendix~\ref{App-B}. The second term is appropriate for the strong superconducting gap, which is not the case here. Then it yields the expression for conductance as
\begin{align}
     G_{sup-PMH}^s-1&\propto-\mathcal{Y}^2\nonumber\\&\times f(K^{-1})T^{2K^{-1}-3}\Big(1+\frac{g(K^{-1})}{T^2}\Big).
\end{align}

One can see that when $K>2/3$, the decrease in temperature increases the resistance value due to superconductivity. This is also in agreement with the relevance of the superconductivity coefficient in the RG concept, as we know that at $K>1/2$ superconductivity enters its relevant regime. In addition, the effect of the Zeeman field on this resistance can be investigated. In this range, the Zeeman gap is relevant. In the presence of attractive interactions, the superconducting gap increases significantly, which greatly reduces the conductance. In this case, the Zeeman gap also increases its strength.

\subsection{Correction to the conductance due to impurity}

Based on the calculations presented in Appendix~\ref{App-C}, we arrived at corrections arising from the forward and backward terms of the charge impurity as
\begin{align}
    R_{imp-c}^{forward}&\propto\mathcal{Y}^2V_c(l)^{2}\nonumber\\&\times f(K^{-1})T^{2K^{-1}-3}\Big(1+\frac{g(K^{-1})}{T^2}\Big),\label{imp-f-correct}\\R_{imp-c}^{backward}&\propto\mathcal{Y}^2V_c(l^*)^{2}\sin^2(2\vartheta_{k_F})\nonumber\\&\times f(K^{-1})T^{2K^{-1}-3}\Big(1+\frac{g(K^{-1})}{T^2}\Big).\label{imp-b-correct}
\end{align}

In addition to the standard backscattering disorder terms, our analysis reveals that forward impurity scattering, typically considered irrelevant for long-range order, introduces a non-negligible correction to long-range correlation functions in the presence of the Zeeman field. This correction is interaction-dependent and emerges from the RG flow equations when disorder and proximity-induced pairing coexist. More interestingly, a mixed term proportional to the product of the impurity and pairing amplitudes appears in the effective action. This term, absent in conventional treatments where these perturbations are studied separately, reflects the nonlinear coupling between charge fluctuations and superconducting order in the one-dimensional helical system. The scaling dimension of this cross-term depends sensitively on the Luttinger parameter and on the strength of the Zeeman gap. In certain interaction regimes, it becomes relevant and modifies the stability of the dominant ordering tendencies, potentially competing with or reinforcing the superconducting or density-wave phases. These results suggest that impurity-pairing interference terms can play a crucial role in determining the phase structure of partial mixed helical superconducting edges. Therefore, a complete phase diagram must account for the cooperative or antagonistic effects of disorder, interactions, and superconductivity under the Zeeman field.

In Eq. (\ref{imp-f-correct}), the impurity coefficient is not renormalized due to $\langle \partial_x\Phi\partial_x\Phi\rangle$ correlations, and the effect of the forward part of the impurity on the super-PMH edge is adjusted by a simple coefficient $V_c(l=0)$. In such a way that at $2/3<K<1$, which in the impurity coefficient has a significant value, Eq. (\ref{imp-f-correct}) has an increasing effect on the correction of resistance due to superconductivity with decreasing temperature. In contrast, the type of correlation introduced by the backward term in Eq. (\ref{imp-b-correct}) requires that the renormalized form of the coefficient $V_c(l^*)$ be used. Therefore, RG-coupled form $R_{imp-c}^{backward}$ using Eq. (\ref{impRG}) is given by
\begin{align}
    R_{imp-c}^{backward}&\propto\mathcal{Y}^2V_c(l=0)^{2}\sin^2(2\vartheta_{k_F})\nonumber\\&\times f(K^{-1})T^{2K+2K^{-1}-5}\Big(1+\frac{g(K^{-1})}{T^2}\Big).\label{imp-b-correct-RG}
\end{align}

In the helical Luttinger liquid framework, the renormalization of the impurity potential provides a crucial mechanism for modifying the low-energy transport behavior. In particular, it contributes to the nontrivial decay of the power-law corrections to conductance. As indicated in Eq. (\ref{imp-b-correct-RG}), when the magnetized Luttinger parameter satisfies $K>1/2$, the system lies in a regime where superconducting pairing is at the onset of relevance, while impurity scattering remains relevant. Under these competing conditions, the impurity potential undergoes renormalization such that, upon lowering the temperature, the correction term to the conductance grows. This enhancement of the impurity-induced correction leads to a suppression of the partially mixed helical edge conductance, thereby highlighting the delicate balance between superconductivity and impurity scattering in determining the fate of low-temperature spin and charge transport.

On the other hand, as the interaction strength approaches the non-interacting limit
$K\rightarrow1$, the RG flow drives the impurity coefficient toward its marginal value. In this regime, impurity backscattering becomes progressively less relevant, leading to a substantial reduction of its impact on carrier substitution and low-energy transport. Consequently, the helical edge channels retain a higher degree of coherence, and the conductance suppression caused by impurity scattering is strongly mitigated. This crossover from a strongly renormalized impurity-dominated regime at $K<1$ toward a nearly ballistic regime at $K\simeq1$ underscores the delicate interplay between electron–electron interactions and impurity effects in helical Luttinger liquids.

In this case, the superconducting gap coefficient benefits significantly from the effective transition the repulsive toward attractive interaction regimes. As the system flows toward $K\rightarrow1$, pairing correlations are enhanced, and the proximity-induced superconducting order becomes more robust against impurity-induced suppression. This regime favors the stabilization of coherent pairwise transport, where the superconducting channel competes less with impurity backscattering and more effectively governs the low-energy conductance. As a result, the helical liquid exhibits a crossover toward a superconductivity-dominated phase, in which the partial suppression of edge transport by impurities is compensated by the emergence of long-range phase-coherent pairing.

The effect of the Zeeman field, among others, is to enhance the stability and effective lifetime of impurity-induced barriers. From the perspective of the Luttinger parameter, an increasing Zeeman field drives the system towards stronger effective interactions, thereby reducing the value of $K$. This shift keeps the system longer within the relevant regime for impurity scattering, namely $1/2<K<1$. As a consequence, impurity backscattering remains more pronounced over an extended range of temperatures and energies, suppressing the conductance more efficiently than in the field-free case. At the same time, this field-induced renormalization alters the competition between impurity and superconducting channels, ultimately modifying the crossover scale between impurity-dominated and superconductivity-dominated transport in the helical liquid. In addition, increasing the magnetic field gap can increase the correction term because of the presence of an additional $\sin^2(2\vartheta_{k_F})$ factor.

Magnetic impurities appear with a spin quantization defect in the helical direction ($x$) and two transverse directions $y$ and $z$. In the first case, the result resistance contains a forward scattering as $\langle\partial_x\Theta \partial_x\Theta\rangle$ thereby $V_{0x}$ is not rescaled. The resistances due to $y$ and $z$ components of magnetic impurity are rescaled by Eq. (\ref{impRG}) resulting in the corrections,
\begin{align}
     R_{imp-y}^{backward}&\propto\mathcal{Y}^2V_{sy}(l=0)^{2}\nonumber\\&\times f(K^{-1})T^{2K+2K^{-1}-5}\Big(1+\frac{g(K^{-1})}{T^2}\Big)\label{imp-y-b-correct-RG},\\R_{imp-z}^{backward}&\propto\mathcal{Y}^2V_{sz}(l=0)^{2}\cos^2(2\vartheta_{k_F})\nonumber\\&\times f(K^{-1})T^{2K+2K^{-1}-5}\Big(1+\frac{g(K^{-1})}{T^2}\Big).\label{imp-z-b-correct-RG}
\end{align}
The power-law damping of the corrections to the conductance due to the transverse impurity spins y and z is similar to that in the long-range charge case. The difference between these three corrections is the factor dependent on the Zeeman field. We note that the transverse y-component of the impurity spin is aligned with the magnetic field. So, the correction due to the y-component of the impurity spin, having no factor, has a maximum value. The z-component of the impurity spin has the factor $\sin^2(2\vartheta_{k_F})$ and the charge impurity has the factor $\cos^2(2\vartheta_{k_F})$. However, increasing the Zeeman gap will tend to increase the charge-correction factor and decrease the z-component correction factor.

\begin{figure}[t!]
    \centering
    \includegraphics[width=8cm]{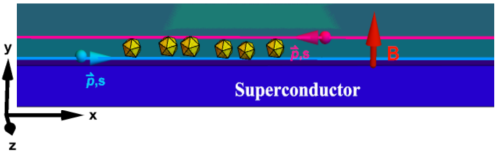}
   \caption{(Color online) Schematic of the superconducting partial mixed helical edge in the presence of many impurities.}
    \label{imp-many}
\end{figure}

 \section{Many impurities}\label{s4}
In helical Luttinger liquids, distinguishing between a single impurity and a finite density of impurities (i.e., disorder) is crucial for understanding the resulting transport properties. A single impurity can be treated within the boundary RG framework, where its relevance or irrelevance is controlled by the Luttinger parameter $K$. As already mentioned above, for $K>1$, the impurity becomes irrelevant at low energies, allowing for nearly ballistic edge transport, whereas for $K<1$, the impurity is relevant and effectively cuts the helical channel, resulting in an insulating fixed point. In contrast, the presence of many impurities introduces random backscattering events and drives the system into a disordered regime. Now, we consider many impurities in the super-PMH edge as shown in Fig. \ref{imp-many}. Remarkably, while a single impurity mainly provides insight into scaling laws and boundary effects, a finite impurity density is more representative of realistic experimental conditions, where the interplay of disorder, Zeeman field, and superconductivity may give rise to rich localization–delocalization crossovers and even phase transitions between topological and trivial regimes.

\subsection{Scaling analysis of impurities in the super-PMH edge}

In the helical Luttinger liquid framework, disorder is introduced by random backscattering terms that couple to the bosonic charge field. Unlike the case of individual impurities, where the perturbation takes the form
$V_i\cos(\sqrt{4\pi}\Phi)$, a random distribution of impurities is modeled by a Gaussian random potential $\xi(x)$, giving rise to the Hamiltonian disorder $\mathcal{H}_{dis}=\int dx \xi(x)\cos(\sqrt{4\pi}\Phi)$,
with the statistical property $(\langle \xi(x)\xi(x') \rangle = \mathcal{D}_\xi \delta(x-x')$. The variance $\mathcal{D}_\xi$ defines the strength of the disorder. Under the RG flow, this parameter obeys the scaling equation,
\begin{equation}
    \frac{d\mathcal{D}(l)}{dl}=(3-2K(l))\mathcal{D}(l),
    \label{disorder-RG}
\end{equation}
where $\mathcal{D}=\frac{2\mathcal{D}_\xi a_0}{\pi v^2}$ and $K$ is the Luttinger parameter renormalized by electron–electron interactions and the external Zeeman field. This scaling relation demonstrates a crucial difference between many impurities and single impurity physics: for $K>3/2$, the disorder are irrelevant, and the helical edge benefits charge-conducting. For $K<3/2$, the disorder become relevant, driving the system toward Anderson localization and suppressing charge transport.

\textbf{The effect of Zeeman Field}: The Zeeman field reduces the effective Luttinger parameter $K$, as spin-charge locking is perturbed and repulsive interactions are effectively enhanced. Consequently, the Zeeman term extends the regime in which the disorder remains relevant ($K<3/2$), thus increasing the likelihood of localization. 

\textbf{Competition with Superconductivity}: In proximity to an s-wave superconductor, an additional perturbation of the form $\Delta \cos(\sqrt{4\pi}\Theta)$ arises, where $\Theta$ is the dual bosonic field associated with spin. This term tends to stabilize the superconducting order by pinning $\Theta$, which competes directly with the disorder-induced pinning of $\Phi$. The resulting phase diagram is determined by the interplay between the scaling of $\mathcal{D}$ and the superconducting coupling $\Delta$.
For the dominant disorder ($\mathcal{D}\rightarrow\infty$), localization suppresses superconductivity.
For sufficiently strong superconducting coupling, long-range coherence overcomes random backscattering, and the system flows toward a superconducting fixed point. Consequently, the Zeeman field, by lowering $K$, biases the system towards the disorder-dominated regime, unless the superconducting proximity is strong enough to offset localization.

\subsection{Correction to conductance of the super-PMH edge}
\label{correct-dis}

We assume the super-PMH edge in the length parametric regime $L_T\ll L_{\Delta}, L_{\mathcal{D}}, L$ where $L_T$, $L_{\Delta}$, and $L_{\mathcal{D}}$ are thermal, superconducting gap, and disorder length scales. Using the memory-function formalism, a well-established approach for interacting conductors~\cite{Gotze1972} and its modern extensions~\cite{Das2016,Bhalla2016}, the low-frequency transport can be calculated. The memory-function framework has been successfully applied to optical and dc conductivities of correlated and gapped systems with phase fluctuations (see e.g. \cite{Kupcic2017,Basov2011}). It is therefore a natural choice to treat impurity-induced relaxation in the proximitized helical edge. In what follows, transport calculations are performed within the low-energy regime, where every relevant energy scale, i.e., the Zeeman gap and the proximity-induced superconducting gap, remains much smaller than the Fermi energy. In this regime, even when $k_BT$ exceeds the smaller gaps, they are only thermally renormalized rather than completely suppressed. So the memory-function formalism remains applicable for transport calculations. Unlike the intrinsic BCS gap, which collapses once $k_BT\sim \Delta_{BCS}$, in our case, the proximate-induced gap in the helical edge acts as an external coupling and therefore survives as a perturbative term even for $k_BT>\Delta$. This thermal softening weakens the proximity effect but simultaneously enhances spin transport by partially releasing the spin degrees of freedom~\cite{Parks2018,Stepniak2015,Valls2010,Kiphart2021}.

On the other hand, the memory-function formalism remains valid in the presence of impurity-induced scattering, since disorder provides the diffusive relaxation mechanism required for current decay, as discussed by Forster \cite{Forster2018} and Das \cite{Das2016}. Similar formulations have been successfully applied to interacting 1D conductors (Rosch and Andrei, \cite{Rosch2000}) and disordered metallic systems~\cite{Gotze1972}. However, here, due to the dimensionless form of energies, even though $\Delta_s > k_BT$, $\mathcal{Y} < k_BT$ always holds.

Employing the conductivity of the long edge~\cite{Mahan,Bakhshipour2024} and using the relation $R=\frac{L}{\sigma}$~\cite{Visuri2020}, we calculate the contribution of isotropic disorder to the correction of the sup-PMH edge conductance as
\begin{equation}
    G_s^L-1\propto-R^L_{edge},
\end{equation}
where $R^L_{edge}=R^L_{sup-PMH}+R^L_{dis}$ with
\begin{align}
    R^L_{super-PMH}&=-\mathcal{Y}^2\frac{L}{a_0}(\frac{2\pi a_0 T}{v})^{2K^{-1}-3},\\R^L_{dis}&=-\mathcal{Y}^2\mathcal{D}(l^*)\frac{L}{a_0}(\frac{2\pi a_0 T}{v})^{2K^{-1}-3}.\label{correct-dis1}
\end{align}
Here, $D(l^*)$ is the renormalized form of disorder for which we consider the temperature to be the largest energy scale. Using flow (\ref{disorder-RG}) the correction of disorder to resistance is found as,
\begin{equation}
R^L_{dis}=-\mathcal{Y}^2\mathcal{D}(l=0)(\frac{2\pi a_0 T}{v})^{2K+2K^{-1}-6}.
\label{correct-dis2}
\end{equation}

For $0.38<K<3/2$, decreasing the temperature increases the correction term of the disorder. By separating this range into repulsive ($0.38<K<1$) and attractive interactions ($1<K<3/2$), a richer interpretation can be achieved. As mentioned above, in the presence of repulsive interactions, the increase of Zeeman field will cause a decrease in $K$ and, as in the case of a single impurity, will become a support for the potential barrier gap. In the attractive regime, applying a stronger Zeeman field effectively increases the Luttinger parameter $K$, which in turn enhances superconducting pairing. 
Physically, this means that the proximity-induced superconducting gap becomes more robust as the system is driven toward stronger attractive interactions. However, in this same parameter range, disorder can remain relevant: Even though superconductivity is strengthened, impurity scattering continues to grow under renormalization and competes with the coherent pairing tendency. This defines an intermediate 'competition window', where both superconductivity and disorder attempt to dominate the low-energy transport properties. Only when the Zeeman field drives the interactions to become sufficiently attractive, beyond a certain threshold, does disorder become irrelevant, and the system flows toward a superconductivity-dominated phase characterized by a stable gap and suppressed impurity effects.

\begin{figure}[t!]
    \centering
    \includegraphics[width=8cm]{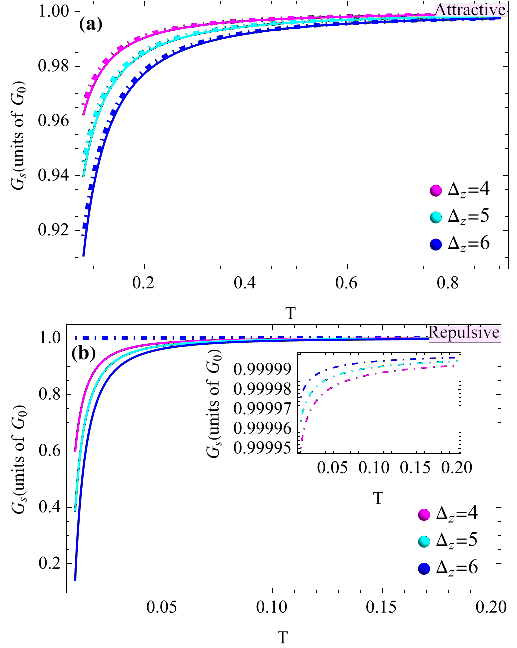}
   \caption{(Color online) Temperature dependence of the conductance in a super-PMH edge for clean (dash–dotted) and disordered (solid) cases in the interaction regimes (a) of attraction and (b) of repulsion, for Zeeman strengths of 4, 5, and 6, a superconductivity gap of 2, and disorder coefficient of 0.002.}
    \label{conductance-dis}
\end{figure}

Figure \ref{conductance-dis} shows the temperature dependence of the spin conductance in the super-PMH edge for both the clean (dash-dotted) and the disordered (solid) cases. Panel (a) corresponds to the attractive interaction regime, where the conductance $G_s$ drops from the quantized value $G_0$ as the temperature decreases, with stronger Zeeman fields $\Delta_z=4,5,6$ leading to enhanced corrections at low $T$. 
It is observed that in this interaction regime, the spin conductance drop behaves almost the same in the absence and presence of disorder, with the difference that considering disorder will cause a further decrease in conductance. Panel (b) illustrates the repulsive regime, in which the spin conductance of the clean edge (dash-dotted) remains nearly quantized, with only small deviations that decay rapidly with temperature, as indicated in the inset. The disordered edge (solid) experiences a significant decrease in conductance. As a result, disorder plays a significant role in the repulsive case, while the attractive case is more robust against disorder and mostly superconductivity is responsible for reducing conductance along with Zeeman's support.

According to our calculations, the $\Phi$ and $\Theta$ fields carry the dominant information for charge and spin, respectively. Then the disorder involving the $\Phi$ field actually directly targets the charge and, by pinning the $\Phi$ field, indirectly affects the spin gap created by theta field pinning, which is contrary to the case of Ref. \cite{Wu2006}. In addition, in our work, the helical state is not the basis, but we are dealing with a PMH edge where the Time Reversal symmetry is broken. Therefore, it is not necessary that the presence of disorder blocks the spin transport by pinning the $\Phi$ field. Here, we consider disorder as a perturbation and, by taking the memory function approach, we find that disorder has a reducing effect on the spin transport, not a complete blocker. The operators $O_1$ and $O_2$ in reference \cite{Wu2006} are the same as the $g_1$ interaction, which in our work does not create a spin gap and is only absorbed in the $g_2$ term. Instead, an evaluation of spin transport in disordered helical edges with broken time-reversal symmetry has been carried out in Ref. \cite{Copenhaver2022}. Our system is based on solving the one-dimensional BHZ model with open boundary conditions~\cite{Qi2011}, which is formed in $HgTe/CdTe$ quantum well heterostructures~\cite{Schmidt2012}.

\section{Density and pairing correlation functions}
\label{s5}
In this section, we investigate the effect of disorder on the charge and spin density wave correlation functions, as well as on superconducting pairing correlations, along the edge of the super-PMH state. To this end, we employ a RG approach to analyze the evolution of the correlation functions under impurity potential operator and evaluate the resulting logarithmic corrections.

\subsection{Relevant super-PMH edge}
\label{s5-1}
We model the system as an effective super-PMH edge such that both the Zeeman and superconducting terms remain relevant, while the operator of disorder resides at its marginal boundary. In this way, we effectively incorporate both Zeeman and superconductivity effects into the helical edge Hamiltonian. Therefore, the Hamiltonian of super-PMH edge in the relevant regimes of superconductivity and Zeeman field takes the quadratic form, 
\begin{equation}
    \mathcal{H}_{sup-PMH}^{relevant}=\frac{\bar{\bar{v}}}{2}\left[\frac{1}{\bar{\bar{K}}}(\partial_x\Phi)^2+\bar{\bar{{K}}}(\partial_x\Theta)^2\right],
    \label{supPMHREV}
\end{equation}
where $\bar{\bar{v}}$ and $\bar{\bar{K}}$ are the new versions of the velocity of excitations and the Luttinger parameter, respectively, given by
\begin{align}
\bar{\bar{K}}&=\sqrt{\frac{\bar{\bar{v}}_F -\frac{g_{fb}}{8\pi}+\frac{g_4}{8\pi}}{\bar{\bar{v}}_F +\frac{g_{fb}}{8\pi}+\frac{g_4}{8\pi}}},
\label{Luttingerparameter-relevant}\\\bar{\bar{v}}&=\sqrt{(\bar{\bar{v}}_F +\frac{g_4}{8\pi})^2-(\frac{g_{fb}}{8\pi})^2}.
\label{Luttingervelocity-relevant}
\end{align}
Here, we have found the renormalized Fermi velocity $\bar{\bar{v}}_F={v}_F\sqrt{1-\frac{(\Delta_s\pm\Delta_z)^2}{\epsilon_F^2}}$.

Note that the effective Hamiltonian in Eq. (\ref{supPMHREV}) is obtained by projecting the Bernevig–Hughes–Zhang model onto the low-energy edge-state subspace, as described in Ref. \cite{Qi2011}. In this approach, the edge dispersion is derived by solving the bulk model with open boundary conditions and retaining only the gapless boundary modes. The proximity-induced superconducting gap and the Zeeman field renormalize the edge-state spectrum, leading to modified Fermi velocity and Fermi momentum. By incorporating these renormalized parameters directly into the projected edge Hamiltonian, we effectively account for the influence of superconductivity and magnetism at the single-particle level. This description is valid in the low-energy regime where the bulk gap of the parent topological insulator remains the largest energy scale and the projection onto the edge-state subspace is justified.

In contrast to the single impurity case, the disorder renormalizes the Luttinger parameter in addition to rescaling itself by $\bar{\bar{K}}$. The RG flow equations can be derived as,
\begin{align}
\frac{d\mathcal{D}(l)}{dl}&=(3-2\bar{\bar{K}}(l))\mathcal{D}(l),
\label{DRG-re-sup-PMH}
\\
\frac{d\bar{\bar{K}}(l)}{dl}&=-\frac{1}{2}\mathcal{D}(l) \bar{\bar{K}}^{2}(l)
\label{KRG-re-sup-PMH}.
\end{align}
In the above equations, using $\bar{\bar{K}}=\frac{1}{2}(3+\bar{\bar{\mathcal{Y}}}_{\vartheta_{k_F}})$ and its expansion to the second order in $\bar{\bar{\mathcal{Y}}}_{\vartheta_{k_F}}$, we get the interaction-dependent RG equations
\begin{align}
\frac{d\mathcal{D}(l)}{dl}&=-\bar{\bar{\mathcal{Y}}}_{\vartheta_{k_F}}(l)\mathcal{D}(l),
\label{DRG-re-sup-PMH-int}
\\
\frac{d\bar{\bar{\mathcal{Y}}}_{\vartheta_{k_F}}(l)}{dl}&=-\frac{9}{4}\mathcal{D}(l)
\label{KRG-re-sup-PMH-int}.
\end{align}
 The interaction dependence of the parameter $\bar{\bar{\mathcal{Y}}}_{\vartheta_{k_F}}$ is defined as $\bar{\bar{\mathcal{Y}}}_{\vartheta_{k_F}}=-1-2\bar{\bar{y}}_{fb}$ where $\bar{\bar{y}}_{fb}=\frac{g_{fb}}{8\pi\bar{\bar{v}}_F}$. An interaction-dependent relevancy analysis for the disorder shows that in the attractive interaction with $\bar{\bar{y}}_{fb}=-1/2$,  the disorder gap coefficient is at its marginal limit. As we move towards weaker attractions $\bar{\bar{y}}_{fb}>-1/2$, this coefficient becomes deeper as we enter its relevant regime. In the absence of interactions $\bar{\bar{y}}_{fb}=0$, the disorder has a significant gap, and as we go towards stronger repulsions $\bar{\bar{y}}_{fb}>0$, it moves toward strong coupling.

\begin{figure}[htbp]
    \centering
    \includegraphics[width=8cm]{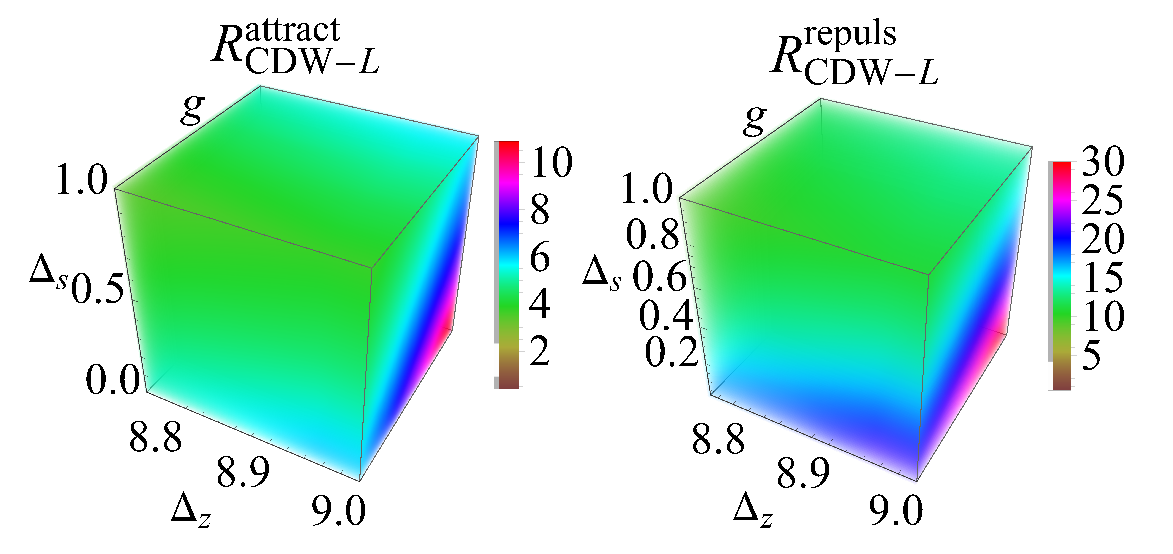}
   \caption{(Color online) Long-range amplitudes of charge density wave
correlation function $R_{\mathrm{CDW}\text{-}L}$ plotted as functions of
the Zeeman field $\Delta_z$, superconducting gap $\Delta_s$, and interaction strength $g$. The color scale represents the magnitude of $R_{\mathrm{CDW}\text{-}L}$. The arrows indicate the direction of
increasing interaction strength. (a) Attractive interaction regime
($g<0$). (b) Repulsive interaction regime ($g>0$).
Parameters are fixed at $x=0.01$ and $\tau=0.1$.
}
    \label{CDW}
\end{figure}

\subsection{Density and pairing correlation functions}
\label{s5-2}
Now, we investigate charge- and spin-density wave correlation functions, as well as spin-singlet (SS) and spin-triplet (TS) superconductivity. Using the standard definitions of these correlation functions \cite{Giamarchi2004,Giamarchi1986,Giamarchi1988,Giamarchi1989,Maier2014}, their explicit expressions with respect to $\mathcal{H}_{sup-PMH}^{relevant}$ can be obtained as (see Appendix \ref{App-D}),
\begin{align}
\mathcal{R}_{CDW}(r)&=\frac{\bar{\bar{K}}F_2(r)}{4\pi^2}+\sin^2(\beta)\sin^2(2\vartheta_{k_F})\frac{F_2(r)}{4\pi^2\bar{\bar{K}}}\nonumber\\&+\frac{\cos^2(\beta)\sin^2(2\vartheta_{k_F})}\cos(2k_Fx){2(\pi a_0)^2}e^{-2\bar{\bar{K}}F_3(r)},
\label{CDWnonperturb}
\end{align}
\begin{align}
\mathcal{R}^{xx}_{SDW}(r)&=-\cos^2(\beta)\frac{\cos^2(2\vartheta_{k_F})}{4\pi^{2}\bar{\bar{K}}}F_2(r)\nonumber\\&+\frac{\sin^2(\beta)\cos^2(2\vartheta_{k_F})\cos(2k_Fx)}{2(\pi a_0)^2}e^{-2\bar{\bar{K}}F_3(r)},\label{xSDWnonperturb}\\
\mathcal{R}^{yy}_{SDW}(r)&=-\frac{\sin^2(2\vartheta_{k_F})}{4\pi^{2}}\bar{\bar{K}}F_2(r)+\sin^2(\beta)\frac{F_2(r)}{4\pi^2\bar{\bar{K}}}\nonumber\\&+\cos^2(\beta)\frac{\cos(2k_F x)}{2(\pi a_0)^2}e^{-2\bar{\bar{K}}F_{3}(r)},\label{ySDWnonperturb}\\ \mathcal{R}^{zz}_{SDW}(r)&=\frac{\cos^2(2\vartheta_{k_F})\cos(2k_F x)}{2(\pi a_0)^2}e^{-2\bar{\bar{K}}F_{3}(r)},\label{zSDWnonperturb}\\
\mathcal{R}^{xy}_{SDW}(r)&=-\mathcal{R}^{yx}_{SDW}(r)\nonumber\\&=-\frac{\sin(2\beta)\cos(2\vartheta_{k_F})\cos(2k_F x)}{4(\pi a_0)^2}e^{-2\bar{\bar{K}}F_{3}(r)},
\label{xySDWnonperturb}\\
\mathcal{R}^{xz}_{SDW}(r)&=-\mathcal{R}^{zx}_{SDW}(r)\nonumber\\&=-\frac{\sin(\beta)\cos^2(2\vartheta_{k_F})\cos(2k_F x)}{2(\pi a_0)^2}e^{-2\bar{\bar{K}}F_{3}(r)},
\label{xzSDWnonperturb}
\\
\mathcal{R}^{yz}_{SDW}(r)&=-\mathcal{R}^{zy}_{SDW}(r)\nonumber\\&=\frac{\cos(\beta)\cos(2\vartheta_{k_F})\cos(2k_F x)}{2(\pi a_0)^2}e^{-2\bar{\bar{K}}F_{3}(r)},
\label{yzSDWnonperturb}\\\mathcal{R}_{SS}(r)&=\frac{\cos^2(2\vartheta_{k_F})}{(2\pi a_0)^2}e^{-2\bar{\bar{K}}^{-1}F_{3}(r)},\label{ss-nonperturb}
\end{align}
\begin{align}\mathcal{R}^{x}_{TS}(r)&=\frac{\cos^2(\beta)}{(2\pi a_0)^2}e^{-2\bar{\bar{K}}^{-1}F_{3}(r)},\label{x-TS-nonperturb}
\\\mathcal{R}^{y}_{TS}(r)&=\frac{\sin^2(\beta)\cos(2\vartheta_{k_F})}{(2\pi a_0)^2}e^{-2\bar{\bar{K}}^{-1}F_{3}(r)},\label{y-TS-nonperturb}
\\\mathcal{R}^{z}_{TS}(r)&=\frac{\cos^2(\beta)\sin^2(2\vartheta_{k_F})}{(2\pi a_0)^2}e^{-2\bar{\bar{K}}^{-1}F_{3}(r)}. \label{z-TS-nonperturb}
\end{align}
where 
\begin{align}
F_2(r)&=\frac{(\bar{\bar{v}}\tau sign(\tau)+a_0)^{2}-x^2}{2[(\bar{\bar{v}}\tau sign(\tau)+a_0)^{2}+x^2]^2},
\label{F2}\\F_{3}(r)&=\frac{1}{2}\ln\Big[\frac{(\bar{\bar{v}}\tau sign(\tau)+a_0)^{2}+x^2}{a_0^{2}}\Big],
\label{F3}\\
\beta&=\tan^{-1}(-\frac{\Delta_s}{\tilde{v}_Fk_F}).
\end{align}

\begin{figure*}[htbp]
    \centering
    \includegraphics[width=16cm]{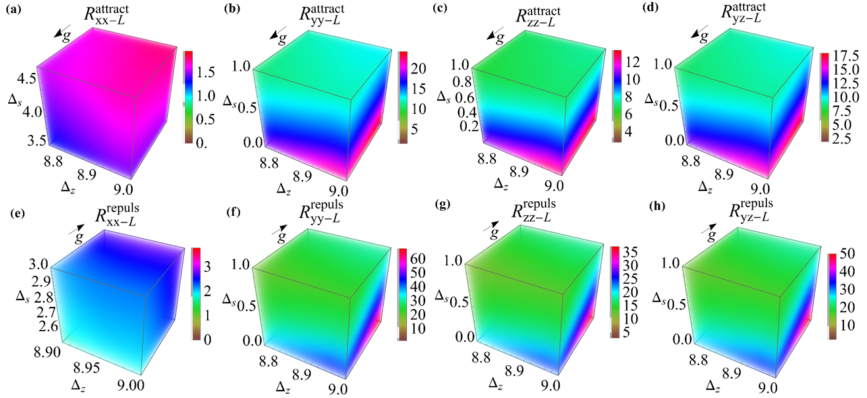}
   \caption{(Color online) Long-range amplitudes of the x-, y-, z-, yz- components of spin density wave correlation functions as functions of the Zeeman field $\Delta_z$, the superconducting gap $\Delta_s$, and the interactions strength $g$. Panels (a) and (e): the x component $R_{xx\text{-}L}$. Panels (b) and (f): the y component $R_{yy\text{-}L}$. Panels (c) and (g): the z component $R_{zz\text{-}L}$. Panels (d) and (h): the yz component $R_{yz\text{-}L}$. The top row (a–d) corresponds to the attractive regime ($g<0$), and the bottom row (e–h) to the repulsive regime ($g>0$). The color scale represents the magnitude of the corresponding spin density wave long-range amplitude. Arrows indicate the direction
of increasing interaction strength. Parameters are fixed at
$x=0.01$ and $\tau=0.1$.}
    \label{SDW}
\end{figure*}

\begin{figure*}[htbp]
    \centering
    \includegraphics[width=16cm]{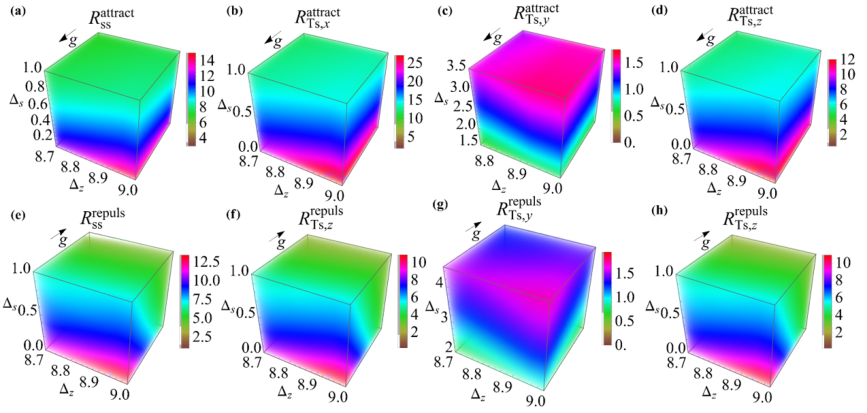}
   \caption{(Color online) Long-range amplitudes of singlet and x-, y-, z-triplet pairing correlation functions as functions of the Zeeman field $\Delta_z$, the superconducting gap $\Delta_s$, and the interactions strength $g$. Panels (a) and (e): the singlet pairing response $R_{ss}$. Panels (b) and (f): the x-triplet pairing response $R_{T_{s,x}}$. Panels (c) and (g): the y-triplet pairing response $R_{T_{s,y}}$. Panels (d) and (h): the z-triplet pairing response $R_{T_{s,z}}$. The color scale represents the magnitude of the corresponding correlation function. The top row (a–d) corresponds to the attractive regime ($g<0$), and the bottom row (e–h) to the repulsive regime ($g>0$). Arrows indicate the direction of increasing interaction strength. Parameters are fixed at $x=0.01$ and $\tau=0.1$.}
    \label{TS}
\end{figure*}

Before discussing the results shown in Figs. \ref{CDW}, \ref{SDW}, and \ref{TS}, we clarify the quantities plotted. In these figures, we present the long-distance amplitudes of the charge-density wave, spin-density wave, and superconducting pairing correlation functions defined in Eqs. (\ref{CDWnonperturb})–(\ref{z-TS-nonperturb}). Specifically, the correlation functions are evaluated at fixed spatial separation $x = 0.01$ and imaginary time $\tau = 0.1$.

To explicitly illustrate the dependence of the long-range charge density wave response (Eq. (\ref{CDWnonperturb})) on the physical parameters, we present in Fig. \ref{CDW} the charge density wave correlation function $R_{\mathrm{CDW}\text{-}L}$ as function of three variables: the Zeeman field strength $\Delta_z$, the superconducting gap $\Delta_s$, and the interaction strength $g$. The color scale represents the magnitude of $R_{\mathrm{CDW}\text{-}L}$, while the arrows indicate the direction of increasing interaction strength. Panel (a) depicts the competition of the Zeeman and superconductivity gaps in the presence of attractive interactions. It is observed that the largest correlation value occurs at the Zeeman maximum and the superconductivity minimum. Increasing the electron attraction reduces the correlation value. In contrast, panel (b) shows the interplay of the two gaps in the repulsive interaction. The largest amount of correlation occurs in the strong Zeeman field and weak superconductivity in the presence of strong repulsive interactions. The correlation lifetime in the repulsive interactions is longer than in the attractive case.

In Fig. \ref{SDW}, we show the long-range amplitudes of the x-, y-, z-, yz- components of spin density wave [Eqs. (\ref{xSDWnonperturb})-(\ref{zSDWnonperturb}) and (\ref{yzSDWnonperturb})] as functions of  the Zeeman field, the superconducting gap, and the interaction. The upper and lower panels correspond to attractive and repulsive interactions, respectively. The color scale represents the magnitude of each corresponding correlation function. Arrows indicate the direction of increasing interaction strength. The $x$ component of the spin-density wave correlation in the presence of attractive [panel (a)] and repulsive [panel (e)] interactions indicates an increasing trend with both superconductivity and Zeeman field. Weaker attractive interactions and stronger repulsive interactions increase this amount sightly. The $y$-phase, in the presence of attractive [panel (b)] and repulsive [panel (f)] interactions, exhibits an increasing behavior with the Zeeman field and a decreasing behavior with superconductivity. Stronger repulsive interactions enhance the correlation, whereas stronger attractive interactions suppress it further. The Zeeman field, superconductivity, and the interactions in the $z$-phase [panels (c) and (g)] have competitive behavior similar to that of the $y$ component, with the difference that the $y$-phase is more stable than the $z$-phase in both regimes. Note that the mixed components $\mathcal{R}^{xy}_{SDW}(r)$ and $\mathcal{R}^{xz}_{SDW}(r)$ have negligible values due to their dependence on the $\sin(\beta)$ or $\sin(2\beta)$ factors. The mixed correlation $\mathcal{R}^{yz}_{SDW}(r)$ exhibits behavior similar to that of the $y$- and $z$-phase correlations in both interaction regimes [see panels (d) and (h)]. Although its magnitude is smaller than that of the dominant y-phase correlation, it is closer in strength to the leading correlation than the other mixed components.

In Fig. \ref{TS}, the long-range amplitudes of the superconducting singlet and x-, y-, z-triplet pairing correlation functions  (Eqs. (\ref{ss-nonperturb})-(\ref{z-TS-nonperturb})) are plotted as functions of $\Delta_z$, $\Delta_s$, and $g$. The color scale represents the magnitude of each corresponding correlation function. The upper and lower panels correspond to attractive and repulsive interactions, respectively. The singlet superconducting phase $\mathcal{R}_{ss}$ in panels (a) and (e) has a maximum correlation value at stronger attractive interactions [panel (a)] and weaker repulsive interactions [panel (e)]. However, the Zeeman gap and the superconducting gap still have an opposite increasing-decreasing effect on the phase. Panels (b) and (f) show the x-triplet pairing correlation. The behavior is similar to that of the singlet case, but the correlation in the triplet is more stable. The $y$ component of pairing correlations increase with increasing superconducting and Zeeman gaps [Panels (c) and (g)]. As the attractive interaction becomes stronger, the correlation reaches a maximum while for repulsive interaction the correlations decreases. The $z$-phase of pairing correlation [Panels (d) and (h)] is similar to the pairing $x$-phase. As a result, among all phases, the dominant phase is the $x$-triplet correlations in the attraction regime.

\subsection{Corrections in the weak disorder regime: Logarithmic corrections}
\label{s5-3}

In this section, we analyze the perturbative effects of impurities on the density and pairing correlation functions in the presence of relevant Zeeman and superconductivity gaps. In low-dimensional systems, even weak disorder can drastically modify the structure of correlation functions. In the clean limit, these functions typically display pure power-law decay, reflecting the critical nature of one-dimensional phases. However, the presence of disorder leads to deviations from this ideal behavior and generates logarithmic corrections. Such corrections originate from multiple scattering of electronic or spin degrees of freedom off local impurities and naturally emerge within perturbative analyses and RG treatments. As such, the decay of correlations is no longer governed by a simple power law but acquires additional logarithmic factors. This indicates the intrinsic sensitivity of one-dimensional systems to disorder. The Hamiltonian of disorder in the basis of super-PMH edge takes the form of $\mathcal{H}_{dis}=\int dx \xi(x)\sin(\beta+2\vartheta_{k_F})\cos(\sqrt{4\pi}\Phi)$.

The logarithmic corrections to the correlation functions induced by disorder can be expressed as (see Appendix \ref{App-E} for details)

\begin{align}
    \mathcal{R}^{log-dis}_{CDW}(r)&=\frac{\cos^2(\beta)\sin^2(2\vartheta_{k_F})\cos(2k_Fx)}{2(\pi a_0)^2}(\frac{a_0}{r})^3L^{dis}_1,
\label{CDW-dis-perturb}\\
\mathcal{R}^{xx,log-dis}_{SDW}(r)&=\frac{\sin^2(\beta)\cos^2(2\vartheta_{k_F})\cos(2k_Fx)}{2(\pi a_0)^2}(\frac{a_0}{r})^3L^{dis}_1,\label{xSDW-dis-perturb}
\\\mathcal{R}^{yy,log-dis}_{SDW}(r)&=\frac{\cos^2(\beta)\cos(2k_F x)}{2(\pi a_0)^2}(\frac{a_0}{r})^3L^{dis}_1,\label{ySDW-dis-perturb}\\ \mathcal{R}^{zz,log-dis}_{SDW}(r)&=\frac{\cos^2(2\vartheta_{k_F})\cos(2k_F x)}{2(\pi a_0)^2}(\frac{a_0}{r})^3L^{dis}_1,\label{zSDW-dis-perturb}\\
\mathcal{R}^{xy,log-dis}_{SDW}(r)&=-\mathcal{R}^{yx,log-dis}_{SDW}(r)\nonumber\\&=-\frac{\sin(2\beta)\cos(2\vartheta_{k_F})\cos(2k_F x)}{4(\pi a_0)^2}(\frac{a_0}{r})^3L^{dis}_1,
\label{xySDW-dis-perturb}\\
\mathcal{R}^{xz,log-dis}_{SDW}(r)&=-\mathcal{R}^{zx,log-dis}_{SDW}(r)\nonumber\\&=-\frac{\sin(\beta)\cos^2(2\vartheta_{k_F})\cos(2k_F x)}{2(\pi a_0)^2}(\frac{a_0}{r})^3L^{dis}_1,
\label{xzSDW-dis-perturb}\\
\mathcal{R}^{yz,log-dis}_{SDW}(r)&=-\mathcal{R}^{zy,log-dis}_{SDW}(r)\nonumber\\&=\frac{\cos(\beta)\cos(2\vartheta_{k_F})\cos(2k_F x)}{2(\pi a_0)^2}(\frac{a_0}{r})^3L^{dis}_1,
\label{yzSDW-dis-perturb}
\end{align}
\begin{align}\mathcal{R}_{SS}^{log-dis}(r)&=\frac{\cos^2(2\vartheta_{k_F})}{(2\pi a_0)^2}(\frac{a_0}{r})^{4/3}L_2^{dis},\label{ss-sup-PMH-log}\\ \mathcal{R}^{x,log-dis}_{TS}(r)&=\frac{\cos^2(\beta)}{(2\pi a_0)^2}(\frac{a_0}{r})^{4/3}L_2^{dis},\label{x-TS-sup-PMH-log}
\\\mathcal{R}^{y,log-dis}_{TS}(r)&=\frac{\sin^2(\beta)\cos(2\vartheta_{k_F})}{(2\pi a_0)^2}(\frac{a_0}{r})^{4/3}L_2^{dis},\label{y-TS-sup-PMH-log}
\\\mathcal{R}^{z,log-dis}_{TS}(r)&=\frac{\cos^2(\beta)\sin^2(2\vartheta_{k_F})}{(2\pi a_0)^2}(\frac{a_0}{r})^{4/3}L_2^{dis},\label{z-TS-sup-PMH-log}
\end{align}
where  
\begin{align}
L^{dis}_1=&\bar{\bar{\mathcal{Y}}}^{-\frac{1}{2}}_{\vartheta_{k_F}}\ln^{-\frac{1}{2}}(\frac{r}{a_0}),\label{L_1-dis}\\L^{dis}_2=&\bar{\bar{\mathcal{Y}}}^{\frac{2}{9}}_{\vartheta_{k_F}}\ln^{\frac{2}{9}}(\frac{r}{a_0}).\label{L_2-dis}
 \end{align}

 In the superconducting PMH edge, the presence of disorder manifests itself in a nontrivial renormalization of correlation functions. This can be compared to the case of quantum wires like~\cite{Gangadharaiah2008,Sun2007} or exploring of magnetic orders in one dimensional conductors~\cite{Braunecker2015}. Within the RG framework, when the disorder operator resides at its marginal boundary in the regime of attractive interactions, the scaling dimensions of the charge- and spin-density wave operators are shifted by logarithmic factors. As a consequence, the power law exponents governing the decay of these correlations acquire logarithmic corrections, leading to a suppression of charge- and spin-density correlations. Note that this effect takes place despite enhancing of the strength of superconducting gap by Zeeman field in the attractive regime. Therefore, while superconducting pairing remains relevant, the logarithmic corrections introduced by the marginal disorder substantially modify the scaling laws of density fluctuations and provide a distinct fingerprint in spin transport along the edge.
In contrast, for pairing correlations, the same disorder operator induces positive logarithmic corrections, which enhance their stability and effectively reinforce the superconducting order.

We emphasize that such weak disorder enhances superconducting pairing correlations refers to the relative stability of pairing channels at the finite distances and energy scales probed by our correlation functions, rather than to an increase of the microscopic proximity gap. Two physical mechanisms underlie this behavior. First, disorder (and single-particle backscattering in particular) disrupts the coherent charge density wave and spin density wave patterns that require long-range phase coherence. Consequently density correlations are suppressed more strongly by disorder. Second, the proximity-induced pairing in our system is an externally imposed, local pairing field whose short-range amplitude is less sensitive to impurity-induced dephasing than the long-wavelength density modes. From the RG perspective, marginal disorder operator generates logarithmic renormalizations of the correlations and suppresses the power-law behavior of density channels via negative logarithmic corrections while making the superconducting pairing correlations more pronounced at the scales considered by positive logarithmic corrections. We stress that this picture applies in the weak-disorder, intermediate-scale regime analyzed in this work; in the opposite limit of very strong disorder or at asymptotically large length scales localization effects eventually dominate and suppress all long-range correlations.



\section {Summary and Conclusions} \label{s6}

In this work, we investigated the response of helical edge spins in a topological superconductor under a Zeeman field in the presence of both single-charge/spin impurities and disorder. In particular, we examined the fate of spin conductance as well as charge- and spin-density correlations and superconducting pairing correlations. For a single impurity, which becomes relevant only in the repulsive regime, the spin conductance is suppressed not only by the superconducting term but also by the additional reduction induced by both charge and spin impurity operators with the y-component of impurity spin leading; this effect is further enhanced with a supporting impurity gap in the repulsions as the Zeeman gap increases. In contrast, in the case of disorder, which reaches its marginal regime under attractive interactions, although conductance still receives a suppression, the Zeeman field simultaneously strengthens the superconducting gap, leading to a competing scenario between disorder and superconductivity. In addition, we find that the dominant phase in the presence of attractive interactions is the x-triplet pairing correlation, and increasing Zeeman enhances it. Moreover, a weak disorder suppresses charge- and spin-density correlations due to logarithmic correction, while at the same time it enhances the stability of both singlet and triplet superconducting phases. Finally, it is interesting to note that this investigation can be extended to richer symmetry classes or exotic bound states \cite{Fleckenstein2019Para,TraversoZiani2020}.

\section*{Data Availability Statement}
The data that support the findings of this study are available from the corresponding author upon reasonable request.

\section*{Acknowledgement}
We are grateful to B. Braunecker, O. Starykh, and N. Traverso Ziani for valuable discussions.

\appendix
\begin{widetext}
\section{Renormalization Group Analysis based on the correlation functions}
\label{App-A}
In this Appendix we outline the derivation of the RG equations used in the main text. Our goal is not to provide an exhaustive treatment, but to clarify the approximations, scaling procedure, and the origin of the flow equations governing the Zeeman-induced PMH edge in the presence of superconducting proximity. We start from the bosonized representation of the partially mixed helical edge, where the Zeeman and proximity-induced terms enter as symmetry-breaking perturbations to the Luttinger liquid fixed point.
\begin{align}
     \mathcal{H}_{sup-PMH}=\frac{v}{2} \left[ \frac{1}{K} (\partial_x \Phi)^2 + K (\partial_x \Theta)^2 \right] +\frac{\Delta}{\pi a_0}\cos\!\big(\sqrt{4\pi}\,\Theta\big),
\end{align}
where in which, first and second terms indicate $\mathcal{H}_{PMH}$ and $\mathcal{H}_{sup}$, respectively.
To arrive at the RG equations, in the first step we calculate the correlation function with respect to the unperturbed state ($\mathcal{H}_{PMH}$), which at $r_1-r_2\gg a_0$ gives the following result
\begin{align}
   \mathcal{R}(r_1-r_2)=\langle e^{ia\sqrt{4\pi}\Theta(r_1)}e^{-ia\sqrt{4\pi}\Theta(r_2)}\rangle_{\mathcal{H}_{PMH}}=e^{-2a^2K^{-1}F_1(r_1-r_2)},
\end{align}
where 
\begin{align}
    F_{1}(r)&=\frac{1}{2}\ln\Big[\frac{(v\tau sign(\tau)+a_0)^{2}+x^2}{a_0^{2}}\Big].
\end{align}
We investigate the correlation function with respect to the full Hamiltonian by expanding to the second order the action term
\begin{align}
   \mathcal{R}(r_1-r_2)=\langle e^{ia\sqrt{4\pi}\Theta(r_1)}e^{-ia\sqrt{4\pi}\Theta(r_2)}\rangle_{\mathcal{H}_{sup-PMH}}.
\end{align}
After some analytical calculations, we arrive at the following expression for the correlation function
\begin{align}
     \mathcal{R}(r_1-r_2)=e^{-2a^2K^{-1}F_1(r_1-r_2)}\Big[1+\frac{\mathcal{Y}^2a^2K^{-1}F_1(r_1-r_2)}{4\pi a_0^4}\int_{a_0}^{\infty}d^2r r^2 e^{-2K^{-1}F_1(r)}\Big],
\end{align}
where $\mathcal{Y}=\frac{4\Delta a_0}{v}$ and $r$ is the relative coordinate of the perturbation term. By reexponenting the above expression, we arrive at the effective $K_{eff}^{-1}$, which allows us to compare the expression with the unperturbed result.
\begin{align}
    K^{-1}_{eff}=K^{-1}-\frac{\mathcal{Y}^2}{4}K^{-2}\int_{a_0}^{\infty} \frac{dr}{a_0}(\frac{r}{a_0})^{3-2K^{-1}}.
    \label{Keff}
\end{align}

One can vary the cutoff ($a_0^\prime=a_0+da_0$)
\begin{align}
    K^{-1}_{eff}=K^{-1}-\frac{\mathcal{Y}^2}{4}K^{-2}\frac{da_0}{a_0}-\frac{\mathcal{Y}^2}{4}K^{-2}\int_{a_0^\prime}^{\infty} \frac{dr}{a_0}(\frac{r}{a_0})^{3-2K^{-1}},
\end{align}
and using the assumption that the exponent should remain unaffected by the cutoff, $K^{-1}$ changes as
\begin{align}
   K^{-1}(a_0^\prime)= K^{-1}(a_0)-\frac{\mathcal{Y}^2(a_0)}{4}K^{-2}(a_0)\frac{da_0}{a_0}.
   \label{Kprime}
\end{align}
Similarly, to get back Eq. (\ref{Keff}) but with $a_0^\prime$ one should rescale the integral and obtain
\begin{align}
    \mathcal{Y}^2(a_0^\prime)=\mathcal{Y}^2(a_0)(\frac{a_0^\prime}{a_0})^{4-2K^{-1}(a_0)}.
    \label{Yprime}
\end{align}
Upon rescaling the original cutoff $a_0$ to the running cutoff $a_0(l)=a_0 e^{l}$ where $l$ denotes the logarithmic length scale that is incremented from $l$ to $l+dl$ and constructing an infinitesimal change in Eqs. (\ref{Kprime}) and (\ref{Yprime}), we obtain the RG flow equations (\ref{YRG}) and (\ref{KRG}). Similarly, using the correlation functions procedure and including the Zeeman and superconducting terms as two sine-Gordon terms in the helical edge states \cite{Bakhshipour2025}, we obtain a pair of RG flows presented in the main text below Eqs. (\ref{YRG}) and (\ref{KRG}). Note that in the present work, we effectively introduced the Zeeman gap in the helical edge. This helped us to properly consider the effects of additional perturbation due to disorder in the presence of a relevant Zeeman that transforms the edge state into a PMH state.

\section{Memory function approach in the presence of impurity}
\label{App-B}
To implement the memory–function method, we start from the full Hamiltonian
\begin{equation}
    \mathcal{H}_{sup-PMH}^{imp}=\mathcal{H}_{PMH}+\mathcal{H}_{sup}+\mathcal{H}_{imp},
\end{equation}
containing
\begin{align}
\mathcal{H}_{PMH}&= \frac{v}{2} \left[ \frac{1}{K} (\partial_x \Phi)^2 + K (\partial_x \Theta)^2 \right],\\\mathcal{H}_{sup} &= \frac{\Delta}{\pi a_0}\cos\!\big(\sqrt{4\pi}\,\Theta\big),\\
\mathcal{H}_{imp-c}&=\int dx V_c\,\delta(x)\big(\frac{1}{\sqrt{4\pi}}\partial_x\Phi(x)+\frac{\sin(2\theta_{k_F})}{\pi a_0}\cos(\sqrt{4\pi}\Phi(x))\big),
\end{align}
where $\Delta=\Delta_s\sin(2\vartheta_{k_F})$ indicates the total superconductivity gap. The low-energy edge theory contains two gap-opening perturbations: The
superconducting term that is gaping out the bosonic field $\Theta$, and the impurity term that locks to $\Phi$. We focus on the spin current operator $j_s= \frac{v}{K} \, \partial_x \Phi$ and evaluate the commutator $F=[j_s,\mathcal{H}_{sup-PMH}^{imp}]$ as
\begin{equation}
F = \left[ \frac{v}{K} (\partial_x \Phi), \frac{\Delta}{\pi a_0} 
\cos(\sqrt{4\pi}\Theta) \right]
= -\frac{iv}{K}  \frac{\Delta}{\pi a_0}
\left[-\sqrt{4\pi}\sin(\sqrt{4\pi}\Theta)\right].
\end{equation}
In the present setup 
$F$ is proportional to the superconducting potential, so
its correlator can be computed with respect to the residual Hamiltonian ($\mathcal{H}_{eff}=\mathcal{H}_{PMH}+\mathcal{H}_{imp}$) perturbatively.

The spin conductivity obtains from the Memory function as
\begin{equation}
\sigma_s(\omega) \;=\;
\frac{\chi_{j_s j_s}}{-\,i\omega + M_s(\omega)}\label{cond},
\end{equation}
where
\begin{equation}
    M_s(\omega,T)\simeq \frac{\langle F;F\rangle^0_{\omega,T}-\langle F;F\rangle^0_{\omega=0,T}}{-\omega \chi(0)},\label{Memory}
\end{equation}
with $\chi(0)=-2vK^{-1}/\pi$. $M_s(\omega,T)$
is built from the retarded correlator evaluated at the impurity location
$\langle F;F\rangle^{eff}_{\omega}$. This construction captures the competition between impurity scattering
and superconducting pairing in the renormalization of the low-frequency spin
response. 

The correlator with respect to $\mathcal{H}_{eff}$ is formed as,
\begin{align}
\langle F;F\rangle_\tau = &\frac{v^2}{K^2}\frac{2\pi\Delta^2}{(\pi a_0)^2}
\left(-\frac{1}{4}\Big\langle T_\tau 
\big( e^{i\sqrt{4\pi}\Theta(x,\tau)} - e^{-i\sqrt{4\pi}\Theta(x,\tau)} \big)
\big( e^{i\sqrt{4\pi}\Theta(0,0)} - e^{-i\sqrt{4\pi}\Theta(0,0)} \big)
\Big\rangle_{\mathcal{H}_{\text{eff}}}\right)\nonumber\\
=& \frac{v^2}{K^2}\frac{2\pi\Delta^2}{(\pi a_0)^2}
\left(-\frac{1}{4}\langle T_\tau (e^{i\sqrt{4\pi}(\Theta(x,\tau)-\Theta(0,0))} 
+ e^{-i\sqrt{4\pi}(\Theta(x,\tau)-\Theta(0,0))})\rangle_{\mathcal{H}_{\text{eff}}}\right).
\end{align}
Using the concept of averaging and expanding the impurity up to the second order, we construct the correlator with respect to the non-perturbative Hamiltonian,
\begin{align}
\langle F;F\rangle_\tau =& \frac{v^2}{K^2}\frac{2\pi\Delta^2}{(\pi a_0)^2}
\left(\frac{e^{i\sqrt{4\pi}(\Theta(x,\tau)-\Theta(0,0))}e^{S_{eff}}}{S_{eff}}\right)\nonumber\\
=& \frac{v^2}{K^2}\frac{2\pi\Delta^2}{(\pi a_0)^2}
\frac{e^{i\sqrt{4\pi}(\Theta(x,\tau)-\Theta(0,0))} e^{-S_{PMH}}\left(1+\tfrac{1}{2}S_{\text{imp}}^2\right)}{e^{-S_{PMH}}}.
\end{align}
Therefore, the result of the main part and the correction one is obtained,
\begin{align}
\langle F;F\rangle_\tau =& \frac{v^2}{K^2}\frac{2\pi\Delta^2}{(\pi a_0)^2}
\Bigg\{ \langle T_\tau e^{i\sqrt{4\pi}(\Theta(x,\tau)-\Theta(0,0))}\rangle_{\mathcal{H}_{PMH}}
+ \frac{V_c^2}{8\pi} \langle T_\tau e^{i\sqrt{4\pi}(\Theta(x,\tau)-\Theta(0,0))}\partial_x\Phi(x=0)\partial_x\Phi(x=0)\rangle_{\mathcal{H}_{PMH}}
\nonumber\\&+\frac{V_c^2}{8(\pi a_0)^2}\sin^2(2\vartheta_{k_F})\langle T_\tau e^{i\sqrt{4\pi}(\Theta(x,\tau)-\Theta(0,0))} e^{i\sqrt{4\pi}(\Phi(x,\tau)-\Phi(0,0))}\rangle_{\mathcal{H}_{PMH}}\Bigg\}.
\label{FF-total}
\end{align}
The first term is the superconductivity correction 
\begin{equation}
\langle F;F\rangle_\tau = \frac{v^2}{K^2}\frac{2\pi\Delta^2}{(\pi a_0)^2}
 e^{-2\pi\langle(\Theta(x,\tau)-\Theta(0,0))^2\rangle},
\end{equation}
where $\langle(\Theta(x,\tau)-\Theta(0,0))^2\rangle=\frac{F_1(x,\tau)}{\pi K}$. Here, 
\begin{equation}
    F_{1}(r)=\frac{1}{2}\ln\Big[\frac{({v}\tau sign(\tau)+a_0)^{2}+x^2}{a_0^{2}}\Big].
\end{equation}
By calculating the correlation at finite temperature and $(x,\tau)\gg a_0$, we arrive at the retarded real-time correlation function. Performing Fourier transform and using the change of variables, one can evaluate integrals and get the ($q,\omega$) dependence correlation as
\begin{align}
    \langle F;F\rangle_{q,\omega}=&\frac{v^2}{K^2}\frac{2\pi\Delta^2}{(\pi a_0)^2}\frac{\sin(\pi K^{-1})a_0^2}{v}(\frac{2\pi a_0 }{\beta v})^{2K^{-1}-2}B\Big(\frac{K^{-1}}{2}-i\frac{\beta v(\frac{\omega}{v}-q)}{4\pi },1-K^{-1}\Big)\nonumber\\
    &B\Big(\frac{K^{-1}}{2}-i\frac{\beta v(\frac{\omega}{v}+q)}{4\pi },1-K^{-1}\Big),
\end{align}
where $B(x,y)$ is the Beta function and $\beta=1/T$ . Plugging the above equation into Eq. (\ref{Memory}), we obtain the expression for the Memory function,
\begin{align}
    M(\omega,T)=&4\pi i K^{-1}\Delta^2 (\frac{2\pi a_0 T}{\beta v})^{2K^{-1}-2}\frac{1}{T}B\Big(\frac{K^{-1}}{2}-i\frac{vq}{4\pi T},1-K^{-1}\Big)B\Big(\frac{K^{-1}}{2}+i\frac{vq}{4\pi T},1-K^{-1}\Big)\nonumber\\&\cot\Big(\pi\frac{K^{-1}}{2}+i\frac{vq}{4T}\Big)\cot\Big(\pi\frac{K^{-1}}{2}-i\frac{vq}{4T}\Big).
\end{align}

To access the conductance, we compute the resistance in the zero length limit of the edge ($L\rightarrow0$) and $T>vq$ by using
\begin{equation}
    R_{imp-c}=\Big[\frac{1}{L}\sum_q \sigma_{s}(q,T)\Big]^{-1}.
\end{equation}
Substituting Eq. (\ref{cond}) into the above relation and converting the summation into the integral yields
\begin{align}
    R_{imp-c}=T^{2K^{-1}-3}f(K^{-1})\Big(1+\frac{g(K^{-1})}{T^2}\Big),
\end{align}
where
\begin{align}
    f(K^{-1})&=\frac{
2^{-2K^{-1}}\, a \, \cot\!\left(\tfrac{K^{-1}\pi}{2}\right) 
\, \Gamma\!\left(\tfrac{1}{2}-\tfrac{K^{-1}}{2}\right)^{2} 
\, \Gamma\!\left(\tfrac{K^{-1}}{2}\right)^{2} 
\sqrt{ \pi^{2} \csc^{2}\!\left(\tfrac{K^{-1}\pi}{2}\right) - 4 \psi^{(1)}\!\left(1-\tfrac{K^{-1}}{2}\right) + 4 \psi^{(1)}\!\left(\tfrac{K^{-1}}{2}\right)}
}
{\pi^{2}\sqrt{-\dfrac{4 e^{iK^{-1}\pi}\pi^{2}}{(-1+e^{iK^{-1}\pi})^{2}} -4\psi^{(1)}\!\left(1-\tfrac{K^{-1}}{2}\right)+4\psi^{(1)}\!\left(\tfrac{K^{-1}}{2}\right)}},
\\g(K^{-1})&=\frac{
2^{-4} v^{2} \pi^{2}\big( e^{iK^{-1}\pi}\pi^{2} + (-1+e^{iK^{-1}\pi})^{2}\psi^{(1)}\!\left(1-\tfrac{K^{-1}}{2}\right) - (-1+e^{iK^{-1}\pi})^{2}\psi^{(1)}\!\left(\tfrac{K^{-1}}{2}\right)\big)}
{3 a^2 (-1+e^{iK^{-1}\pi})^{2}}.
\end{align}
Here, $a$, $\Gamma$, and $\psi^{(1)}$ indicate the lattice constant, the Gamma function, and the first order of PolyGamma function ($\text{PolyGamma}[1,x]$), respectively. The correction terms are obtained in Appendix \ref{App-C}.
\section{correction of the conductance arisen by the single impurity}
\label{App-C}
The second and third terms in Eq. (\ref{FF-total}), are contributions of forward and backward of the charge impurity to correct conductance of edge, respectively. Because the impurity is a perturbative term and gap out edge, the correlation $\langle \partial_x \Phi(x=0)\partial_x\Phi(x=0)\rangle$ and $e^{-2\pi \langle (\Phi(x,\tau)-\Phi(0,0))^2\rangle}$ are of order 1. Expression of correction given by
\begin{align}
    \langle F;F\rangle^{forward}_{\tau,correct}=\frac{v^2\Delta^2(l)}{4K^2(\pi a_0)^2}V_c^2(l),\\
    \langle F;F\rangle^{backward}_{\tau,correct}=\frac{v^2\Delta^2(l)}{4K^2(\pi a_0)^2}\sin^2(2\vartheta_{k_F})V_c^2(l^*).
\end{align}
Here, the impurity in the first term does not renormalized. According to
\begin{align}
   \partial_{x_1} \partial_{x_2} \langle \Phi(r_1) \Phi(r_2) \rangle=\frac{(v\tau sign(\tau)+a_0)^{2}-x^2}{2[(v\tau sign(\tau)+a_0)^{2}+x^2]^2},
\end{align}
We obtain the scaling dimension of thecorrelation function,
\begin{equation}
    \partial_{x} \partial_{x} \langle \Phi(r) \Phi(0) \rangle \sim \frac{\tau^2}{\tau^4}=\frac{1}{\tau^2},
\end{equation}
and get to $L^{(2-2)}$. But the gap coefficient with scaling dimension $L^{(2-K)}$ in backward term corresponds to Eq. (\ref{impRG}) is renormalized,
\begin{equation}
    V_c(l^*)=V_c(l=0)(\frac{v}{a_0T})^{1-K}.
\end{equation}
Therefore, the correct terms are obtained as,
\begin{align}
    R_{imp-c}^{forward}&\propto\mathcal{Y}^2V_c(l)^{2}f(K^{-1})T^{2K^{-1}-3}\Big(1+\frac{g(K^{-1})}{T^2}\Big),\label{imp-f-correct-app}\\R_{imp-c}^{backward}&\propto\mathcal{Y}^2V_c(l=0)^{2}sin^2(2\vartheta_{k_F})f(K^{-1})T^{2K+2K^{-1}-5}\Big(1+\frac{g(K^{-1})}{T^2}\Big).\label{imp-b-correct-app}
\end{align}
\section{Correlation functions of the clean $super-PMH$ edge}
\label{App-D}
In this Appendix, we calculate the charge-density, spin-density, and superconducting pairing operators in the basis of the super-PMH edge, i.e., a helical edge partially spin-mixed due to superconducting proximity. The fermionic field operators can be rewritten as 
\begin{align}
\psi^\prime_r
= \sum_{\sigma=\pm}\psi^\prime_{r\sigma}
= \sum_{\sigma=\pm}
\langle \mathbf{\zeta}_r(k_F) | \sigma \rangle
\, e^{i r k_F x}\, \psi_r ,
\end{align}
where $r=R,L$ and $\psi_{R/L}$ are the chiral annihilation operators. In addition, $\mathbf{\zeta}_r(k_F)$ are the eigen spinors of the BdG super-PMH edge given by
\begin{align}
    \mathbf{\zeta}_R&=\cos(\beta/2)e^{-i\pi/4}\chi_R+\sin(\beta/2)e^{i\pi/4}\chi_L,
    \\\mathbf{\zeta}_L&=-\sin(\beta/2)e^{-i\pi/4}\chi_R+\cos(\beta/2)e^{i\pi/4}\chi_L.
\end{align}
Here, the $\chi_{R/L}$ are the eigen spinors of PMH edge in absence of superconductivity (as introduced in Sec. \ref{s2}).
Therefore, one can obtain the field operators of the system as 
\begin{align}
    \psi^\prime_R&=\Big[\sqrt{2}\cos(\beta/2)e^{i\pi/4}\cos(\vartheta_{k_F})-i\sqrt{2}\sin(\beta/2)e^{-i\pi/4}\sin(\vartheta_{k_F})\Big]e^{ik_Fx}\psi_R,
    \\\psi^\prime_L&=\Big[-\sqrt{2}\sin(\beta/2)e^{i\pi/4}\cos(\vartheta_{k_F})-i\sqrt{2}\cos(\beta/2)e^{-i\pi/4}\sin(\vartheta_{k_F})\Big]e^{-ik_Fx}\psi_L.
\end{align}
The effects of the Zeeman and superconducting gaps enter respectively through the phase parameters $\vartheta_{k_F}=\frac{1}{2}\tan^{-1}(-\frac{\Delta_z}{k_Fv_F}), \beta=\tan^{-1}(-\frac{\Delta_s}{k_F\tilde{v}_F})$, and, more generally, through the renormalized Luttinger parameters, which encode the combined influence of both gaps on the scaling dimensions of the corresponding operators. Plugging the above fields into the standard definitions  of charge-density, spin-density, and paring operators yields   
\begin{align}
    \mathcal{O}_{CDW}&=(\psi^\dagger_R\psi_R+\psi^\dagger_L\psi_L)-\sin(\beta)\sin(2\vartheta_{k_F})(\psi^\dagger_R\psi_R-\psi^\dagger_L\psi_L)\nonumber\\&-\cos(\beta)\sin(2\vartheta_{k_F})(e^{-2ik_Fx}\psi^\dagger_R\psi_L
    +e^{2ik_Fx}\psi^\dagger_L\psi_R),
    \\\mathcal{O}^x_{SDW}&=\cos(\beta)\cos(2\vartheta_{k_F})(\psi^\dagger_R\psi_R-\psi^\dagger_L\psi_L)-\sin(\beta)\cos(2\vartheta_{k_F})(e^{-2ik_Fx}\psi^\dagger_R\psi_L
    +e^{2ik_Fx}\psi^\dagger_L\psi_R),
    \\\mathcal{O}^y_{SDW}&=\sin(2\vartheta_{k_F})(\psi^\dagger_R\psi_R
    +\psi^\dagger_L\psi_L)-\sin(\beta)(\psi^\dagger_R\psi_R
    -\psi^\dagger_L\psi_L)-\cos(\beta)(e^{-2ik_Fx}\psi^\dagger_R\psi_L
    +e^{2ik_Fx}\psi^\dagger_L\psi_R),
    \\\mathcal{O}^z_{SDW}&=i\cos(2\vartheta_{k_F})(e^{-2ik_Fx}\psi^\dagger_R\psi_L
    -e^{2ik_Fx}\psi^\dagger_L\psi_R),
    \\\mathcal{O}_{ss}&=\cos(2\vartheta_{k_F})\psi^\dagger_R\psi^\dagger_L,
    \\\mathcal{O}^x_{TS}&=-\cos(\beta)\psi^\dagger_R\psi^\dagger_L,
    \\\mathcal{O}^y_{TS}&=\sin(\beta)\cos(2\vartheta_{k_F})\psi^\dagger_R\psi^\dagger_L,
    \\\mathcal{O}^z_{TS}&=-i\cos(\beta)\sin(2\vartheta_{k_F})\psi^\dagger_R\psi^\dagger_L.
\end{align}
The bosonized form of the above operators can be obtained as
\begin{align}
    \mathcal{O}_{CDW}&=\frac{1}{\sqrt{4\pi}}\partial_x\Phi(r)-\sin(\beta)\sin(2\vartheta_{k_F})\frac{1}{\sqrt{4\pi}}\partial_x\Theta(r)-\cos(\beta)\sin(2\vartheta_{k_F})\frac{1}{\pi a_0}\sin(\sqrt{4\pi}\Phi(r)+2k_Fx),
    \\\mathcal{O}^x_{SDW}&=\cos(\beta)\cos(2\vartheta_{k_F})\frac{1}{\sqrt{4\pi}}\partial_x\Theta(r)-\sin(\beta)\cos(2\vartheta_{k_F})\frac{1}{\pi a_0}\sin(\sqrt{4\pi}\Phi(r)+2k_Fx),
    \\\mathcal{O}^y_{SDW}&=\sin(2\vartheta_{k_F})\frac{1}{\sqrt{4\pi}}\partial_x\Phi(r)-\sin(\beta)\frac{1}{\sqrt{4\pi}}\partial_x\Theta(r)-\cos(\beta)\frac{1}{\pi a_0}\sin(\sqrt{4\pi}\Phi(r)+2k_Fx),
    \\\mathcal{O}^z_{SDW}&=-\cos(2\vartheta_{k_F})\frac{1}{\pi a_0}\cos(\sqrt{4\pi}\Phi(r)+2k_Fx),
    \\\mathcal{O}_{ss}&=\frac{i}{2\pi a_0}\cos(2\vartheta_{k_F})e^{-i\sqrt{4\pi}\Theta(r)},
    \\\mathcal{O}^x_{TS}&=-\frac{i}{2\pi a_0}\cos(\beta)e^{-i\sqrt{4\pi}\Theta(r)},
    \\\mathcal{O}^y_{TS}&=\frac{i}{2\pi a_0}\sin(\beta)\cos(2\vartheta_{k_F})e^{-i\sqrt{4\pi}\Theta(r)},
    \\\mathcal{O}^z_{TS}&=\frac{1}{2\pi a_0}\cos(\beta)\sin(2\vartheta_{k_F})e^{-i\sqrt{4\pi}\Theta(r)}.
\end{align}
Substituting the above bosonized form of the operators into the formula of the correlation function~\cite{Giamarchi2004}
\begin{align}
\mathcal{R}_{CDW}(r_1, r_2) &= \langle T_\tau \mathcal{O}_{CDW}(r_1) \mathcal{O}_{CDW}(r_2) \rangle, \label{RCDW} \\
\mathcal{R}^{ij}_{SDW}(r_1, r_2) &= \langle T_\tau \mathcal{O}^i_{SDW}(r_1) \mathcal{O}^j_{SDW}(r_2) \rangle, \\
\mathcal{R}_{SS}(r_1, r_2) &= \langle T_\tau \mathcal{O}_{SS}(r_1) \mathcal{O}_{SS}(r_2) \rangle, \\
\mathcal{R}^{ij}_{TS}(r_1, r_2) &= \langle T_\tau \mathcal{O}^i_{TS}(r_1) \mathcal{O}^j_{TS}(r_2) \rangle, \label{RTS}
\end{align}
gets the corresponding expressions for the correlation functions [Eqs. (\ref{CDWnonperturb})-(\ref{x-TS-nonperturb})]. Here $T_\tau$ is the imaginary-time ordering operator, $r_1 = (x_1, v \tau_1)$, $r_2 = (x_2, v \tau_2)$, and $i,j = x,y,z$. We set $r_1 = r$ and $r_2 = 0$ henceforward.

\section{Logarithmic correction with the marginal disorder operator}
\label{App-E}
For completeness, we present here the detailed derivation of the logarithmic corrections to the correlation functions induced by disorder. While the main text discusses the qualitative impact of impurity scattering, the explicit steps of the perturbative expansion and the renormalization group analysis are given below. These calculations show how disorder, although absent at first order, contributes at higher orders and generates logarithmic terms that modify the simple power-law decay of correlations in one-dimensional systems. We perform the procedure for the spin density wave correlation function in the y direction, and the remaining correlations can be performed according to this section. Considering the correlation with respect to $\mathcal{H}^{dis}_{sup-PMH}$, we enter the disorder as a perturbation
\begin{align}
    R_{SDW}^{yy,log-dis}&=\langle\cos(\sqrt{4\pi} \Phi(x_1,\tau_1+2k_Fx_1))\cos(\sqrt{4\pi} \Phi(x_2,\tau_2+2k_Fx_2))\rangle_{\mathcal{H}_{sup-PMH}^{dis}}\\&=\cos(\sqrt{4\pi} \Phi(x_1,\tau_1+2k_Fx_1))\cos(\sqrt{4\pi} \Phi(x_2,\tau_2+2k_Fx_2))e^{-S_{sup-PMH}^{relevant}-S_{dis}},
\end{align}
where $S_{sup-PMH}^{relevant}$ and $S_{dis}$ are the actions of the Hamiltonian of the super-PMH edge and the perturbative disorder Hamiltonian, respectively. By expanding the disorder term up to the second order, we enter the contribution of disorder in correlation function and construct correlator that is averaging regard to super-PMH quadratic action,
\begin{align}
    R_{SDW}^{yy,log-dis}&=\frac{1}{8}\Big(\frac{V_{dis}^{\frac{1}{2}}\sin(\beta+2\vartheta_{k_F})}{2\pi\bar{\bar{v}} a_0}\Big)^2\sum_{\epsilon_1=\pm1}\sum_{\epsilon_2=\pm1}\nonumber\\&\int dx_3 d\tau_3 dx_4 d\tau_4\langle e^{i\sqrt{4\pi}\epsilon_1\Phi(x_1,\tau_1)+2k_fx_1}e^{-i\sqrt{4\pi}\epsilon_1\Phi(x_2,\tau_2)+2k_Fx_2}e^{i\sqrt{4\pi}\epsilon_2\Phi(x_3,\tau_3)}e^{-i\sqrt{4\pi}\epsilon_2\Phi(x_4,\tau_4)}\rangle.
\end{align}
Computing the correlations leads to the following result,
\begin{align}
    R_{SDW}^{yy,log-dis}&=\frac{\cos(2k_F(x_1-x_2))}{2}e^{-2KF_1(r_1-r_2)}+\frac{1}{4}\Big(\frac{V_{dis}^{\frac{1}{2}}\sin(\beta+2\vartheta_{k_F})}{2\pi \bar{\bar{v}} a_0}\Big)^2\cos(2k_F(x_1-x_2))\sum_{\epsilon_1=\pm1}\sum_{\epsilon_2=\pm1}\nonumber\\&\int dx_3 d\tau_3 dx_4 d\tau_4 \delta(x_3-x_4)e^{-2\bar{\bar{K}}F_1(r_1-r_2)}e^{2\bar{\bar{K}}F_1(r_3-r_4)}e^{2\epsilon_2\bar{\bar{K}}[F_1(r_1-r_3)-F_1(r_1-r_4)+F_1(r_2-r_4)-F_1(r_2-r_3)]}.
\end{align}

By multiplying the non-perturbative factor in the above expression, we arrive at the appropriate function for applying the transformations.
\begin{align}
    \bar{R}_{SDW}^{yy,log-dis}&=1+\frac{\pi}{2}\Big(\frac{V_{dis}^{\frac{1}{2}}\sin(\beta+2\vartheta_{k_F})}{\pi\bar{\bar{v}} a_0}\Big)^2\bar{\bar{K}}^2F_1(r_1-r_2)\int_{r^\prime>a_0}dr^\prime r^{\prime2}e^{2\bar{\bar{K}}F_1(r^\prime)}\nonumber\\&=1+\frac{1}{2}\Big(\frac{V_{dis}a_0\sin^2(\beta+2\vartheta_{k_F})}{\pi \bar{\bar{v}}^2}\Big) \bar{\bar{K}}^2\ln\Big(\frac{(r_1-r_2)}{a_0}\Big)\Big\{dl+\int_{r^\prime>a^\prime_0}\frac{dr^\prime}{a^\prime_0} (\frac{r^{\prime2}}{a^\prime_0})^{2-2\bar{\bar{K}}}\Big\},
\end{align}

\begin{align}
    \bar{R}_{SDW}^{yy,log-dis}=e^{\int_0^{l_r}\frac{\mathcal{D}\bar{\bar{K}}^2}{2}\ln(\frac{r}{a_0})dl},
\end{align}
where $\mathcal{D}=\frac{2V_{dis}a_0\sin^2(\beta+2\vartheta_{k_F})}{\pi v^2}$. By putting the coefficient $\frac{\cos(2k_F(x_1-x_2))}{2}e^{-2\bar{\bar{K}}F_1(r)}$ back in place and by performing this procedure on other correlations, we get to the following

\begin{align}
    R_{SDW}^{yy,log-dis}\propto e^{-2\bar{\bar{K}}\ln(\frac{r}{a_0})+\int_0^{l_r}\frac{9\mathcal{D}}{8}\ln(\frac{r}{a_0})dl},\\R_{ss}\propto e^{-2\bar{\bar{K}}^{-1}\ln(\frac{r}{a_0})+\int_0^{l_r}\frac{\mathcal{D}}{2}\ln(\frac{r}{a_0})dl},
\end{align}
where we must consider the marginal boundary of disorder gap in $\bar{\bar{K}}=\frac{3}{2}$. Using Eq. (\ref{KRG-re-sup-PMH-int}) and $\bar{\bar{K}}=\frac{1}{2}(3+\bar{\bar{\mathcal{Y}}}_{\vartheta_{k_F}})$ and some calculations, we find correlation functions as follows

\begin{align}
    R_{SDW}^{yy,log-dis}\propto(\frac{a_0}{r})^3L_1^{dis},\\R_{ss}\propto(\frac{a_0}{r})^{\frac{4}{3}}L_2^{dis},
\end{align}
where 
\begin{align}
    L_1^{dis}=\exp[-\frac{1}{2}\int_0^{l_r}\bar{\bar{\mathcal{Y}}}_{\vartheta_{k_F}}(l)dl],\label{L1}\\L_2^{dis}=\exp[\frac{2}{9}\int_0^{l_r}\bar{\bar{\mathcal{Y}}}_{\vartheta_{k_F}}(l)dl]\label{L2}.
\end{align}
In the separatrix between the relevant and irrelevant regimes, we have $\bar{\bar{\mathcal{Y}}}_{\vartheta_{k_F}}=\frac{3}{2}\mathcal{D}^{\frac{1}{2}}$. By inserting last equation in Eq. (\ref{KRG-re-sup-PMH-int}) and integrating, one can obtain
\begin{equation}
    \bar{\bar{\mathcal{Y}}}_{\vartheta_{k_F}}(l)=\frac{\bar{\bar{\mathcal{Y}}}_{\vartheta_{k_F}}}{1+\bar{\bar{\mathcal{Y}}}_{\vartheta_{k_F}}l}.\label{separatrix}
\end{equation}
We substitute Eq. (\ref{separatrix}) in Eqs. (\ref{L1}) and (\ref{L2}) and perform integration. Then, taking into account the limit $r\rightarrow\infty$, in which $\bar{\bar{\mathcal{Y}}}_{\vartheta_{k_F}}(l_r)\sim \frac{1}{l_r}$, the coefficients $L_1^{dis}$ and $L_2^{dis}$ are obtained as 
\begin{align}
L_1^{dis}=\bar{\bar{\mathcal{Y}}}_{\vartheta_{k_F}}^{-\frac{1}{2}}\ln^{-\frac{1}{2}}(\frac{r}{a_0}),\\L_2^{dis}=\bar{\bar{\mathcal{Y}}}_{\vartheta_{k_F}}^{\frac{2}{9}}\ln^{\frac{2}{9}}(\frac{r}{a_0}),
\end{align}
which are the same as Eqs. (\ref{L_1-dis}) and (\ref{L_2-dis}) in the main text.

\end{widetext}


\end{document}